\documentclass[finalold]{agujournal2019}
\usepackage{url} 
\usepackage{lineno}
\usepackage[inline]{trackchanges} 
\usepackage{soul}
\usepackage{amsmath}
\usepackage{amssymb}
\usepackage{amsthm}
\usepackage{mathrsfs}

\usepackage{xcolor}
\definecolor{darkblue}{rgb}{0,0,.5}
\usepackage[colorlinks=true,
            linkcolor=darkblue, urlcolor=darkblue, citecolor=darkblue,
            raiselinks=true,
            bookmarks=true,
            bookmarksopenlevel=1,
            bookmarksopen=true,
            bookmarksnumbered=true,
            hyperindex=true,
            plainpages=false,
            pdfpagelabels=true,
            pdfstartview=FitH,
            pdfstartpage=1,
            pdfpagelayout=OneColumn]{hyperref}
\usepackage{mathtools}
\DeclarePairedDelimiter{\diagfences}{(}{)}
\newcommand{\diag}{\operatorname{diag}\diagfences}

\usepackage{verbatim}

\usepackage{multirow}

\usepackage{natbib}

\usepackage{threeparttable}

\usepackage{csquotes}
\usepackage{array}
\newcolumntype{P}[1]{>{\arraybackslash}p{#1}}
\newcolumntype{M}[1]{>{\centering\arraybackslash}m{#1}}

\usepackage[inline]{trackchanges}
\addeditor{ZN}

\usepackage{todonotes}

\draftfalse

\journalname{JGR: Solid Earth}

\begin{document}

\title{A Discontinuous Galerkin Method for Simulating 3D Seismic Wave Propagation in Nonlinear Rock Models: Verification and Application to the 2015 M$_w$ 7.8 Gorkha Earthquake}

\authors{Zihua Niu\affil{1}, Alice-Agnes Gabriel\affil{2,1}, Sebastian Wolf\affil{3}, Thomas Ulrich\affil{1}, Vladimir Lyakhovsky\affil{4}, Heiner Igel\affil{1}}

\affiliation{1}{Department of Earth and Environmental Sciences, Ludwig-Maximilians-Universit\"at M\"unchen, Munich, Germany}
\affiliation{2}{Scripps Institution of Oceanography, UC San Diego, La Jolla, CA, USA}
\affiliation{3}{Technical University of Munich, Munich, Germany}
\affiliation{4}{Geological Survey of Israel, Jerusalem 9692100, Israel}


\begin{keypoints}
\item We propose and verify a 3D discontinuous Galerkin method for nonlinear seismic wave propagation on high-performance computing systems.
\item The 2015 M$_w$ 7.8 Gorkha earthquake simulations show co-seismic wave speed reductions from $<$0.01\% to $>$50\%, varying with fault slip and geology.
\item The nonlinear model captures low-frequency ground motion amplification in soft sediments, highlighting key effects for seismic hazard analysis.
\end{keypoints}

\begin{abstract} 
The nonlinear mechanical responses of rocks and soils to seismic waves play an important role in earthquake physics, influencing ground motion from source to site. Continuous geophysical monitoring, such as ambient noise interferometry, has revealed co-seismic wave speed reductions extending tens of kilometers from earthquake sources. 
However, the mechanisms governing these changes remain challenging to model, especially at regional scales.
Using a nonlinear damage model constrained by laboratory experiments, we develop and apply an open-source 3D discontinuous Galerkin method to simulate regional co-seismic wave speed changes during the 2015 M$_w$7.8 Gorkha earthquake. 
We find pronounced spatial variations of co-seismic wave speed reduction, ranging from $<$0.01\% to $>$50\%, particularly close to the source and within the Kathmandu Basin. The most significant reduction occurs within the sedimentary basin and varies with basin depths, while wave speed reductions correlate with the fault slip distribution near the source.
By comparing ground motions from simulations with elastic, viscoelastic, elastoplastic, and nonlinear damage rheologies, we demonstrate that the nonlinear damage model effectively captures low-frequency ground motion amplification due to strain-dependent wave speed reductions in soft sediments. 
We verify the accuracy of our approach through comparisons with analytical solutions and assess its scalability on high-performance computing systems. The model shows near-linear strong and weak scaling up to 2048 nodes, enabling efficient large-scale simulations. 
Our findings provide a physics-based framework to quantify nonlinear earthquake effects and emphasize the importance of damage-induced wave speed variations for seismic hazard assessment and ground motion predictions.
\end{abstract}

\section*{Plain Language Summary}
Earthquakes cause significant changes in the mechanical properties of rocks and soils, including reductions in seismic wave speeds. These changes, recorded over the past two decades using advanced monitoring techniques, such as ambient noise analysis, reveal valuable information about underground conditions. However, existing models cannot fully capture the complex nonlinear behavior of rocks and soils during an earthquake from source to site. To address this, we extend SeisSol, an open-source software for simulating seismic waves, to model 3D nonlinear wave propagation. 
We demonstrate the efficient execution of the code on powerful computers.
This enhancement allows us to study co-seismic wave speed changes while accounting for complex fault geometry and surface topography. We apply this tool to the 2015 M$_w$ Gorkha, Nepal, earthquake and find significant variations in wave speed reductions, ranging from less than 0.01\% to over 50\%, with the largest reductions concentrated in sedimentary basins. Comparisons with other models demonstrate that the nonlinear damage model employed in this study effectively captures the amplification of low-frequency ground motions by soft sediments, a key factor in understanding earthquake impacts. These insights improve our ability to assess seismic hazards and guide the design of infrastructure better equipped to withstand earthquakes.

\section{Introduction}
\label{sec:intro}
Large earthquakes generate strong ground motions that pose a significant threat to civil structures and human life \citep{ben2022grand}. 
Physics-based models of rocks and soils are essential for simulating potential ground motions from earthquakes in numerical simulations that can account for the spatial heterogeneity and complex surface topography of the Earth's lithosphere \citep{cui2010scalable,taufiqurrahman2022broadband,roten2023implementation}.
Linear models have successfully explained key phenomena in seismic wave propagation, such as wave field amplification in soft sediments \citep{moczo1993wave,van2022amplification}, directivity effects of large earthquakes \citep{boatwright1982analysis,roten2014expected,wollherr2019landers}, and resonance in near-surface structures, including surface topography \citep{lee2009effects,hartzell2014ground} and sedimentary basins \citep{castellaro2023resonance}.

In recent decades, nonlinear mechanical responses of rocks to seismic waves have been widely observed, covering distances from a few kilometers to over one hundred kilometers from the source \citep{sens2006passive,gassenmeier2016field,lu2022regional}.
Temporal variations in seismic wave speeds during and after earthquakes have been observed using techniques such as repeating earthquakes  \citep{poupinet1984monitoring,bokelmann2000evidence,schaff2004coseismic}, cross-correlation of the ambient noise or aftershock recordings between seismic station pairs \citep{sens2006passive,brenguier2008towards,qiu2020temporal}, and auto-correlation of data at individual stations \citep{bonilla2019monitoring,qin2020imaging,li2023daily}. 
In these observations, rocks typically exhibit a rapid co-seismic reduction in seismic wave speeds, followed by long-term recovery \citep{gassenmeier2016field}. Measured magnitudes of such co-seismic wave speed reduction range from less than 1\% up to over 10\%, depending on factors such as rock type, distance from the source, depth of interests, and the temporal resolution of the monitoring technique \citep{brenguier2014mapping,wang2021near}. Notably, auto-correlation analyses at single stations reveal that co-seismic reductions in wave speed up to 8\% are possible at depths between 1~km and 3~km within 20 minutes after an earthquake \citep{bonilla2021detailed}. Co-seismic wave speed changes under dynamic perturbation are sensitive to rheology, ambient stress, and thermal and hydraulic conditions \citep{manogharan2022experimental,lu2022regional}. Such changes are potentially new observables that can be extracted from seismic waves to probe subsurface structure and rheology. However, observations of co-seismic wave speed changes may not be adequately captured by linear elastic or visco-elastic models \citep{johnson2005slow,riviere2015set,manogharan2022experimental}, indicating the need for more advanced physics-based frameworks.

The nonlinear mechanical responses become most prominent when seismic waves propagate through soft sediments, typically located a few hundred meters below the ground surface \citep{wang2021near}. Soft sediments typically exhibit low seismic wave speeds, amplifying the strain field to values exceeding 10$^{-3}$ and reducing the shear modulus by more than 50\% \citep{roten20123d,van2022amplification}.  This behavior is accompanied by the damping of ground motion amplitudes \citep{rajaure2017characterizing} and a change in the frequency components of seismograms toward lower values \citep{bonilla2011nonlinear,castro2020new}. Accounting for such nonlinear mechanical responses is crucial for modeling ground motions at both low frequencies \citep[$\le$1 Hz, ][]{roten2014expected} and high frequencies \citep{roten2016high}.

Capturing co-seismic wave speed changes relies on adequate nonlinear rock models. Some of such nonlinear models originate from thermodynamic processes at the microscopic scale \citep{iwan1967class,delsanto2003modeling,lebedev2014unified}. These models usually introduce more parameters than those constrained by observations \citep{wang2021near}. As a practical compromise, continuum damage mechanics (CDM) models are based on simplified assumptions about microscopic material deficiencies and describe macroscopic stress-strain relationships using fewer parameters \citep{kachanov1986introduction,desmorat2016anisotropic,gabriel2021unified}. Within this framework, the CDM model by \citet{lyakhovsky1997distributed} and the internal variable model (IVM) by \citep{berjamin2017nonlinear} have been shown to reproduce laboratory measurements of co-seismic wave speed changes in rocks \citep{renaud2012revealing,feng2018short,manogharan2022experimental,niu2024modeling}. For unconsolidated sediments, such as soil, the loss of stiffness under cyclic loading is effectively described by a hyperbolic shear modulus reduction curve \citep{kramer2024geotechnical,vardanega2013stiffness}.
 
Previous studies have developed numerical methods for modeling co-seismic wave speed changes in 1D \citep{remillieux2017propagation,berjamin2017nonlinear} and 2D \citep{berjamin2019plane, niu2024modeling}, which have been validated through laboratory experiments. The fourth-order staggered-grid finite difference method, implemented in the software AWP-ODC, resolves shear modulus reduction using the IWAN model \citep{iwan1967class} in 3D, with a focus on capturing nonlinear effects in soft sediments for ground motion simulations \citep{cui2010scalable,roten2023implementation}. Consolidated rocks, such as granite, also experience co-seismic wave speed reductions \citep{shokouhi2017slow}, which remain mostly smaller than 1\%. Resolving such small changes is computationally expensive using the IWAN model \citep{roten2023implementation}. Leveraging this phenomenon as a probe for rock types and subsurface physical conditions \citep{riviere2015set,manogharan2022experimental} requires the development of a numerical framework capable of resolving 3D co-seismic wave speed changes in consolidated rocks. Such a framework would act as a critical bridge, enabling realistic regional-scale modeling of co-seismic wave speed changes directly informed by laboratory data. However, to the best of the authors' knowledge, this approach remains unrealized to date.

To fill this gap, we here propose and validate a novel algorithm based on the discontinuous Galerkin method \citep{cockburn1989tvbA,dumbser2006arbitrary,dumbser2008unified} for modeling seismic wave propagation in 3D nonlinear rock rheologies. We implement this algorithm in the open-source software SeisSol \citep{heinecke_petascale_2014,uphoff_extreme_2017,krenz20213d,uphoff_2024_14051105}, which is specifically suited for field-scale seismic wave propagation simulations involving heterogeneous velocity models and complex geometries.
We verify the implementation by comparison against analytical solutions and present scaling tests on the Frontera supercomputer \citep{stanzione_frontera_2020}.

Using this framework, we simulate co-seismic wave speed changes and ground motions during the 2015 $M_W$ 7.8 Gorkha earthquake in the Kathmandu Valley. 
This earthquake occurred directly beneath the Kathmandu Valley \citep{fan2015detailed},
causing over 9,000 fatalities, extensive property damage, and significant loss of life in Nepal. 
Ground motion records reveal that the Kathmandu basin experienced unexpectedly weak high-frequency motions but larger low-frequency motions compared to empirical predictions \citep{takai2016strong}. 
This behavior has been attributed to nonlinear site response \citep{castro2020new}.
To evaluate this hypothesis, we utilize an experimentally constrained nonlinear model, IVM, to simulate the co-seismic wave speed changes in rocks \citep{niu2024modeling}. We also adapt IVM such that it captures the hyperbolic shear modulus reduction curve in soft sediments. By integrating laboratory data, our simulation results quantify the spatial variability of field-scale co-seismic wave speed changes and their impact on peak ground motions, offering important insights for seismic hazard assessment.

\section{Methods}
\label{sec:method}
When nonlinear rock rheology is incorporated into seismic wave propagation simulations, the governing wave equations are classified as nonlinear hyperbolic partial differential equations \citep[PDE,][]{lax2005weak}. 
A key characteristic of these equations is their potential for solutions to develop spatial discontinuities, even if the initial conditions are smooth \citep{leveque2002finite}. Solving these equations requires an algorithm that can adequately resolve discontinuities while maintaining numerical stability. Additionally, to allow realistic large-scale earthquake simulations and energy efficiency, the implementation must scale efficiently across a large number of compute ranks \citep{carrington2008high,cui2010scalable,heinecke2014petascale,ilsche2019power,uphoff2020flexible,krenz20213d}.

This section describes how we formulate the two nonlinear damage rock models employed in this work as a system of nonlinear hyperbolic PDEs.
We then outline the spatial and temporal discretization of these PDEs using the discontinuous Galerkin method \citep{hesthaven2007nodal,cockburn2012discontinuous}.

\subsection{Mathematical framework for nonlinear wave propagation in damaged rocks}
\label{subsec:mathematical_damage_models}

To model co-seismic wave speed changes and their impact on ground motions, we adopt the recent mathematical framework by \citet{niu2024modeling} that utilizes a continuum damage model \citep[CDM,][]{lyakhovsky1997distributed} and an internal variable model \citep[IVM,][]{berjamin2017nonlinear}. Both models have been shown to quantitatively match laboratory data \citep{manogharan2022experimental,feng2018short,niu2024modeling}. 2D solutions for co-seismic wave speed changes modeled with the IVM implemented in the DG method have been validated against the results of the finite volume method \citep{niu2024modeling}. 

In the following, we present a unified DG algorithm for nonlinear wave propagation, designed to accommodate any nonlinear rock model explicitly formulated as a system of hyperbolic equations, including IVM and CDM. This approach extends our previous 2D implementation of IVM to 3D and applies our 3D discontinuous Galerkin (DG) method to model wave propagation using the CDM nonlinear rock model.

Hyperbolic PDEs are required for implementation in SeisSol \citep{uphoff_2024_14051105}. Previous work implemented linear visco-elasticity \citep{kaser2007arbitrary,uphoff2020flexible} and Drucker-Prager elasto-plasticity \citep{wollherr2018off} using the DG algorithm for linear hyperbolic equations. 
In contrast, CDM and IVM introduce nonlinear hyperbolic PDEs, which we summarize as follows: 


\begin{equation}
    \begin{cases}
      \dfrac{\partial \varepsilon_{ij}}{\partial t} 
      &= \dfrac{1}{2} \left( \dfrac{\partial v_i}{\partial x_j} + \dfrac{\partial v_j}{\partial x_i} \right) \\
      \rho \dfrac{\partial v_i}{\partial t} 
      &= \dfrac{\partial \sigma_{ij} (\underset{=}{\varepsilon},\alpha) }{\partial x_j} \\
      \dfrac{\partial \alpha}{\partial t} 
      &= r_{\alpha}  (\underset{=}{\varepsilon},\alpha)
    \end{cases},
    \label{nonlinear damage models}
\end{equation}
where $\underset{=}{\varepsilon} = \varepsilon_{ij}$ and $\sigma_{ij}$ denote, respectively, the total strain and stress tensors, $v_i$ is the vector for particle velocity, and $\rho$ is the material mass density. $\alpha$ is a damage variable, which is 0 for intact rock and 1 for fully damaged rock. $r_{\alpha}$ defines the evolution rate of the damage variable $\alpha$ as a function of the strain tensor and the damage variable itself.

IVM and CDM are both extensions of the classical linear elastic stress-strain relationship that is parameterized with two Lam\'e parameters, i.e., $\lambda_0$ and $\mu_0$ \citep{landau1986theory}. The differences between the two models lie in how they are extended to include nonlinear functions of the stress tensor $\sigma_{ij} (\underset{=}{\varepsilon},\alpha)$, and how the source term $r_{\alpha}  (\underset{=}{\varepsilon},\alpha)$ is defined.

For the IVM \citep{berjamin2017nonlinear}, we write
\begin{equation}
  \begin{cases}
      \sigma_{ij} (\underset{=}{\varepsilon},\alpha) = (1 - \alpha) 
      (\lambda_0 I_1 \delta_{ij} + 2 \mu_0 \varepsilon_{ij}
      + \sigma^{\text{mur}}_{ij})  \\
      r_{\alpha}  (\underset{=}{\varepsilon},\alpha) = \dfrac{1}{\gamma_b \tau_b} [ \dfrac{1}{2} \lambda_0 I_1^2 + \mu_0 I_2 + W^{\text{mur}} - \phi(\alpha) ]\\
    \end{cases},
    \label{ivm-model}
\end{equation}
where $\phi(\alpha) = \gamma_b [\alpha/(1-\alpha)]^2$ is the storage energy, $\gamma_b$ is the scale of $\phi(\alpha)$ with units in pascals (Pa), and $\tau_b$ is the time scale of damage evolution. $I_1 = \varepsilon_{kk}$ and $I_2 = \varepsilon_{ij} \varepsilon_{ij}$ are two strain invariants.

The original IVM framework can incorporate the classical Murnaghan nonlinear elasticity \citep{murnaghan1937finite} with three additional material parameters $l_0$, $m_0$, and $n_0$ to account for third-order terms in the non-quadratic components of the elastic energy function $W^{\text{mur}} = (l-m)/3 I_1^3 + m I_1 I_2 + n I_3$, where $I_3 = \delta_{ijk}\varepsilon_{i1} \varepsilon_{j2} \varepsilon_{k3}$. This leads to the additional stress component $\sigma^{\text{mur}}_{ij} = a_0 \delta_{ij} + a_1 \varepsilon_{ij} + a_2 \varepsilon_{ik} \varepsilon_{kj}$, where the coefficients $a_0 = l_0 I_1^2 - (m_0 - 1/2n_0)(I_1^2 - I_2)$, $a_1 = (2m_0 - n_0)I_1$, and $a_2 = n$. $\delta_{ijk}$ denotes the Levi-Civita permutation symbol.

While Murnaghan nonlinear elasticity is useful for modeling some instances of stress-induced anisotropy \citep{sharma2010wave}, it may not adequately explain the observed co-seismic wave speed reductions under dynamic stress fields \citep{gassenmeier2016field,berjamin2017nonlinear,niu2024modeling}. 
Therefore, in the following, we choose to set $l_0 = m_0 = n_0 = 0$ to exclude the additional terms of Murnaghan nonlinear elasticity in our proposed algorithm. 
This also ensures that $\sigma^{\text{mur}}_{ij} = W^{\text{mur}} = 0$ in Eq. (\ref{ivm-model}). 
However, in Sections \ref{subsec:veri_riemann} and \ref{subsec:veri_frequency}, we demonstrate that our proposed algorithm remains generic and can accurately resolve nonlinear effects resulting from a simplified Murnaghan nonlinear elasticity in 1D.

For the CDM \citep{lyakhovsky1997distributed,lyakhovsky2016dynamic}, we write
\begin{equation}
  \begin{cases}
      \sigma_{ij} (\underset{=}{\varepsilon},\alpha) = \lambda_0 I_1 \delta_{ij} - \alpha \gamma_r \sqrt{I_2} \delta_{ij} + [2 (\mu_0 + \alpha \xi_0 \gamma_r) - \alpha \gamma_r \xi] \varepsilon_{ij}\\
      r_{\alpha}  (\underset{=}{\varepsilon},\alpha) = 
      \begin{cases}
      C_d \gamma_r I_2 (\xi - \xi_0) &\text{, if $\xi - \xi_0 >$ 0}\\
      0 &\text{, if $\xi - \xi_0 \leq$ 0}
      \end{cases}
    \end{cases},
    \label{cdb-model}
\end{equation}
where $\gamma_r$ is a third modulus originating from the homogenization of parallel cracks \citep{lyakhovsky1997non}, and $C_d$ is a damage evolution coefficient. $\xi = I_1/\sqrt{I_2}$ is derived from the two strain invariants. It grows from $-\sqrt{3}$ for isotropic compression to $\sqrt{3}$ for isotropic extension. The damage $\alpha$ starts to accumulate as the strain state deviates farther enough from the isotropic compression. This is expressed as $\xi - \xi_0 > 0$, where $\xi_0$ is a material parameter that is usually negative for rocks \citep{lyakhovsky2016dynamic}.

In this work, we propose a generic algorithm that can be used for either IVM or CDM.
Both models can generally be formulated as a nonlinear hyperbolic system of conservation laws with an additional source term following \citet{dumbser2008unified}:

\begin{align}
    \dfrac{\partial u_p}{\partial t} + \dfrac{\partial F_p^d (\underset{-}{v},\underset{=}{\varepsilon},\alpha)}{\partial x_d} = s_p(\underset{-}{v},\underset{=}{\varepsilon},\alpha)
    \label{nonlinear hyper pdes},
\end{align}
where $u_p = (\varepsilon_{xx},\varepsilon_{yy},\varepsilon_{zz},\varepsilon_{xy},\varepsilon_{yz},\varepsilon_{zx},v_x,v_y,v_z, \alpha)^T$ is a vector of the conservative variables. $\varepsilon_{xx}$, $\varepsilon_{yy}$, $\varepsilon_{zz}$, $\varepsilon_{xy}$, $\varepsilon_{yz}$, and $\varepsilon_{zx}$ are six components of the strain tensor $\underset{=}{\varepsilon} = \varepsilon_{ij}$; $v_x$, $v_y$ and $v_z$ are the three components of the particle velocity vector $\underset{-}{v}$. The flux term $F_p^d$ represents the rates at which the conservative variable $u_p$ gets transferred through a unit area in the direction $x_d$ \citep{leveque2002finite}. The source vector $s_p = (0,0,0,0,0,0,0,0,0, r_{\alpha})^T$ with only one non-zero element $r_{\alpha}$ defined in Eq. (\ref{ivm-model}) for IVM or Eq. (\ref{cdb-model}) for CDM.

\subsection{Numerical discretization of the nonlinear wave equations}
Our implementation adopts the Arbitrary-accuracy DERivative (ADER) discretization in time \citep{titarev2002ader,dumbser2008unified,gassner2011explicit}, and the discontinuous Galerkin (DG) discretization in space \citep{cockburn1989tvbA,dumbser2008unified}. Here, we apply a linearization to the nonlinear hyperbolic PDEs to simplify the adaptation of the algorithm to both damage models, as outlined in Section \ref{subsec:mathematical_damage_models}. This linearization also minimizes the necessary changes to the existing data structure in SeisSol \citep{uphoff2020flexible,uphoff_2024_14051105}. We provide a detailed description of the method in this section and \ref{sec:app_dg_algorithm} and will demonstrate in Section \ref{subsec:veri_riemann} that the algorithm still converges using linearization.

We subdivide the computational domain into tetrahedral elements. Within each element $\mathcal{T}_m$, we use a modal discontinuous Galerkin approach to approximate the conservative variables as $\underset{-}{u} \approx \underset{-}{u}^h$, employing Dubiner’s orthogonal polynomial basis functions, $\phi_l(\underset{-}{x})$ \citep{cockburn2012discontinuous}. 
The temporal evolution of the solution is captured using time-dependent coefficients $Q_{lp}(t)$ defined as:
\begin{align}
    u^h_k(\underset{-}{x},t) = \sum_{l=1}^L U_{lk}(t) \phi_l(\underset{-}{x})\text{, $k$ = 1, 2, ..., $K$}
    \label{modal dg space},
\end{align}
where the index $l$ runs from 1 to $L = (p + 1)(p + 2)(p + 3)/6$ for a polynomial degree $p$. The index $k$ runs from 1 to $K$, the number of elements in the conservative variables $u_p$ in Eq. (\ref{nonlinear hyper pdes}). We discretize the time-dependent coefficients using the ADER scheme with a Taylor series as 
\begin{align}
    U_{lp}(t) = \sum_{i=0}^{N} \dfrac{(t-t_n)^i}{i!} \mathcal{D}_{lp}^i
    \label{eq:taylor_expansion},
\end{align}
where $\mathcal{D}_{lp}^0 = U_{lp}(t_n)$, and $\mathcal{D}_{lp}^i = \left.\dfrac{\partial^i U_{lp}}{\partial t^i} \right|_{t=t_n}$ for $i \ge 1$. 

This discretized system is solved in two steps. First, we linearize the nonlinear hyperbolic system and estimate $\mathcal{D}_{lp}^i$ using the Cauchy-Kovalevskaya approach \citep{kovalevskaja1874theorie}. In the following, we refer to this step as the \enquote{prediction step}. It allows us to obtain the estimated $U_{lp}(t)$ within one stage, as opposed to the Runge-Kutta method \citep{butcher2007runge,gassner2011explicit}. In the second step, we use the predicted $U_{lp}(t)$ to integrate the conservative variables over time while adequately addressing spatial discontinuities at element interfaces, which we refer to as the \enquote{correction step}. In \ref{sec:app_dg_algorithm}, we detail the algorithm to solve these discretized nonlinear wave equations proposed in this work, including how we implement free-surface and absorbing boundary conditions.

\section{Verification against analytical solutions}
\label{sec:verification}

In this section, we verify the proposed numerical algorithm by solving three problems with known analytical solutions. It is essential to confirm that the proposed numerical scheme converges to the correct solutions before applying it to large-scale seismological applications, for which it is impossible to derive analytical solutions for nonlinear wave equations in 3D. 

We first compare our numerical solutions for plane waves in 3D with two existing analytical solutions in 1D: (1) the nonlinear Riemann problem and (2) the generation of high-frequency harmonics from a single-frequency source. For 3D analysis, we show that the proposed algorithm can accurately resolve stress-induced anisotropy of CDM, in agreement with the analytical solutions from \citet{hamiel2009brittle}.

\subsection{The nonlinear 1D Riemann problem}
\label{subsec:veri_riemann}

The Riemann problem is a canonical benchmark with analytical solutions for nonlinear hyperbolic PDEs in one dimension \citep{leveque2002finite}. It is defined by initial conditions with a single discontinuous interface, where the variables have one set of uniform values on one side of the interface while having another set of different uniform values on the other side. The Riemann problem is widely used to assess whether numerical algorithms can accurately resolve discontinuities in solutions, which is an important feature of nonlinear hyperbolic PDEs.

We use a plane shear wave in 3D to configure the 1D Riemann problem. The plane shear wave comprises $\varepsilon_{xy}$ and $v_y$. We set the remaining components to zero. We define the wavefront as parallel to the $y-z$ plane, such that the domain only varies in the $x$ direction, which simplifies Eqs. (\ref{nonlinear damage models}) to:

\begin{equation}
    \begin{cases}
      \dfrac{\partial \varepsilon_{xy}}{\partial t} 
      &= \dfrac{1}{2} ( \dfrac{\partial v_y}{\partial x} ) \\
      \rho \dfrac{\partial v_y}{\partial t} 
      &= \dfrac{\partial \sigma_{xy} (\varepsilon_{xy}) }{\partial x}
    \end{cases},
    \label{1d Riemann eqs}
\end{equation}
where we define $\sigma_{xy} = 2 \mu (1 - \beta \varepsilon_{xy}) \varepsilon_{xy}$ as a nonlinear function of $\varepsilon_{xy}$ with $\beta$ being the first order nonlinear coefficient \citep{landau1986theory}.

This formulation is comparable to a 1D reduction of Murnaghan nonlinear elasticity, as described after Eq. (\ref{ivm-model}).
\citet{meurer2002wave} provide analytical solutions to the Riemann problem for Eqs. (\ref{1d Riemann eqs}), incorporating the simplified 1D nonlinear stress-strain relationship.

We choose material parameters and initial conditions to show the accuracy of our proposed algorithm for materials with strong nonlinearity. Therefore, we set the following initial conditions for the Riemann problem.

\begin{equation}
    [\varepsilon_{xy}, v_y]^T = 
    \begin{cases}
      [0.1, -0.5]^T \text{ for $x<$ 0} \\
      [0.2, -1.0]^T \text{ for $x\ge$ 0}
    \end{cases}.
    \label{1d Riemann ICs}
\end{equation}

\begin{figure}[hptb]
    \centering
    \includegraphics[width=1\columnwidth]{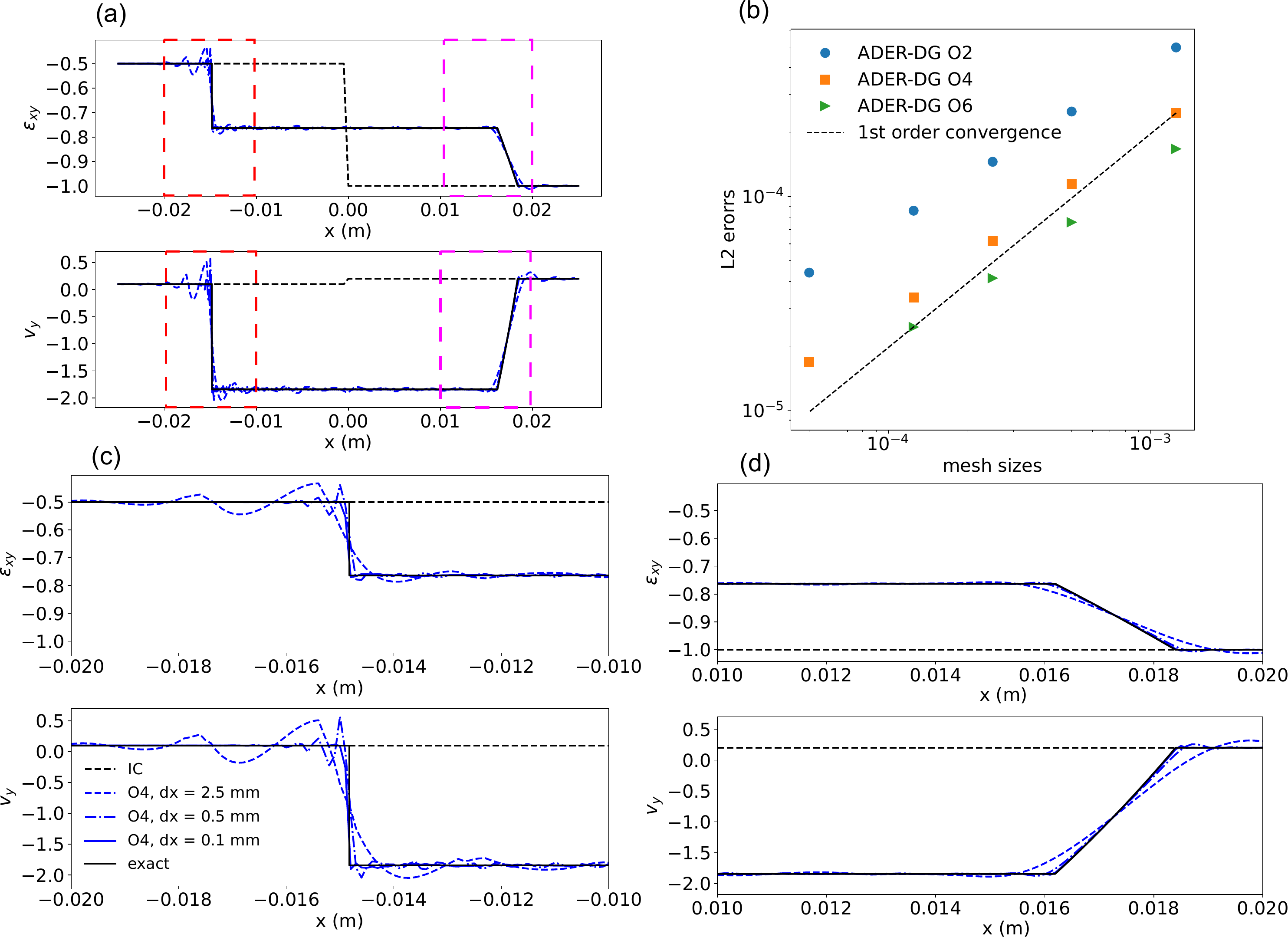}
    \caption{\small Comparison of the analytical and the numerical solutions with varying mesh resolution $h$ and polynomial degrees $p$ for the Riemann problem. (a) Comparison of numerical and analytical solutions of $v_y$ and $\varepsilon_{xy}$ using shape functions of polynomial degree 3 (O4, representing convergence rate of order 4). We show solutions for three mesh sizes: $h$ = 2.5~mm (dashed blue curves), $h$ = 0.5~mm (dash-dotted blue curves) and $h$ = 0.1~mm (solid blue curves). The initial conditions (IC) are illustrated as dashed black curves, and the analytical solutions are given in solid black curves.
    (b)  Convergence analysis showing the error decay with decreasing mesh size $h$, for simulations using basis functions of polynomial degrees 1 (O2, blue dots), 3 (O4, orange rectangles), and 5 (O6, green triangles). The dashed black line indicates first-order convergence as a reference.
    Panels (c) and (d) highlight specific features of (a): the shock wavefront (inside the dashed red rectangles) in (c) and the rarefaction wavefront (inside the dashed pink rectangles) in (d).}
    \label{fig:riemann_problem}
\end{figure}

These initial conditions are also shown as dashed curves in Fig. \ref{fig:riemann_problem}. We set $\rho = 1.0$, $\mu = 1.0$ and $\beta = 10.0$. The black curves shown in Fig. \ref{fig:riemann_problem} are the corresponding analytical solutions evaluated after 4~ms. The solutions feature one shock wave (interface with sharp discontinuities, marked with red dashed rectangles) and one rarefaction wave (a smooth transition from one state on the left to another state on the right, highlighted by purple rectangles). 

We compare this analytical solution to several numerical results obtained with a polynomial order $p=3$ on three mesh sizes: $h$ = 2.5~mm (dashed blue curves), $h$ = 0.5~mm (dash-dotted blue curves) and $h$ = 0.1~mm (solid blue curves). Figs.  \ref{fig:riemann_problem}c and \ref{fig:riemann_problem}d focus on the numerical solutions at the shock wavefront and at the rarefaction wavefront. 
The shock wave exhibits stronger spatial oscillations than the rarefaction wave, primarily due to solution variations within each element.  
The amplitude and wavelength of these oscillations both decrease as the mesh is refined,  indicating that oscillations can be effectively suppressed with mesh refinement. 

We analyze the convergence rates for different orders of polynomial basis functions and present the results in Fig. \ref{fig:riemann_problem}b. We quantify the $L_2$ errors in our numerical simulations at $t$ = 4 ms using the $L_2$ norms of the differences between the analytical solution $u^{ana}$ and the numerical solutions $u^{num}$. We determine the convergence rate by analyzing the reduction of $L_2$ errors with mesh size $h$ on a logarithmic scale.
The observed convergence rate remains first order across all polynomial degrees tested (1 to 5), indicating that this algorithm does not achieve arbitrarily high-order accuracy at discontinuities. 
Nonetheless, we still observe lower $L_2$ errors with higher-order basis functions on the same mesh \citep[p-convergence,][]{wollherr2018off}. We will discuss the underlying causes and potential improvements in Section \ref{subsec:discussion_accuracy_and_efficiency}.

\newpage

\subsection{1D frequency modulation by nonlinear materials}
\label{subsec:veri_frequency}

The generation of harmonics from a single-frequency source is a mathematically intriguing problem in nonlinear wave propagation. It is widely used to quantify material nonlinearity in acoustic testing and non-destructive evaluation \citep{shah2009non,matlack2015review,jiao2025nondestructive}. This behavior is a distinctive and general feature of wave propagation in nonlinear materials, existing in both the Murnaghan nonlinear elasticity and the nonlinear stress-strain relationship in Eq. (\ref{cdb-model}) of CDM.

For the 1D Murnaghan nonlinear elasticity defined in Eq. (\ref{1d Riemann eqs}), we use the 1D analytical asymptotic solutions from \citet{mccall1994theoretical} derived using perturbation theory, which describes how the amplitudes of generated harmonics depend on the nonlinear parameters of the material, the propagation distance, and the source amplitude. We use this analytical reference solution in the following to show that our proposed algorithm can accurately resolve the generation of harmonics in 1D nonlinear numerical simulations, exemplarily for 1D Murnaghan nonlinear elasticity. 

We adopt the same plane shear wave description as in Section \ref{subsec:veri_riemann} for the single-frequency source setup and solve the same nonlinear wave equations as in Eqs. (\ref{1d Riemann eqs}). The simulation is carried out in a cubic domain [-0.025, 0.025] m $\times$ [-0.025, 0.025] m $\times$ [-0.025, 0.025] m, with periodic boundary conditions on all faces. We define the initial conditions for the plane wave such that the wavelength is 0.05 m, matching the length of the simulation domain:

\begin{equation}
    [\varepsilon_{xy}, v_y]^T = 
    [V_0/c_s, V_0]^T \times \sin{(2\pi k x)},
    \label{1d sine ICs}
\end{equation}
where $k$ = 20 m$^{-1}$ and $c_s = \sqrt{\mu/\rho}$ is the shear wave speed. We set $\mu = 82.7$ GPa, $\rho = 2473$ kg/m$^3$, and vary the wave amplitude $V_0$ and the nonlinear coefficient $\beta$ to assess whether the simulation results can quantitatively match the analytical asymptotic solutions at a small propagation distance in Eq. (34) of \citet{mccall1994theoretical}. 
We note that the shear modulus defined here is unrealistically high for rocks; however, these parameters are chosen solely to verify that the numerical solutions are mathematically consistent with the asymptotic solutions. Additionally, the asymptotic solution from \citet{mccall1994theoretical} indicates that the amplitude of the second-order harmonics does not depend on $\mu$.

\begin{figure}[hptb]
    \centering
    \includegraphics[width=0.8\columnwidth]{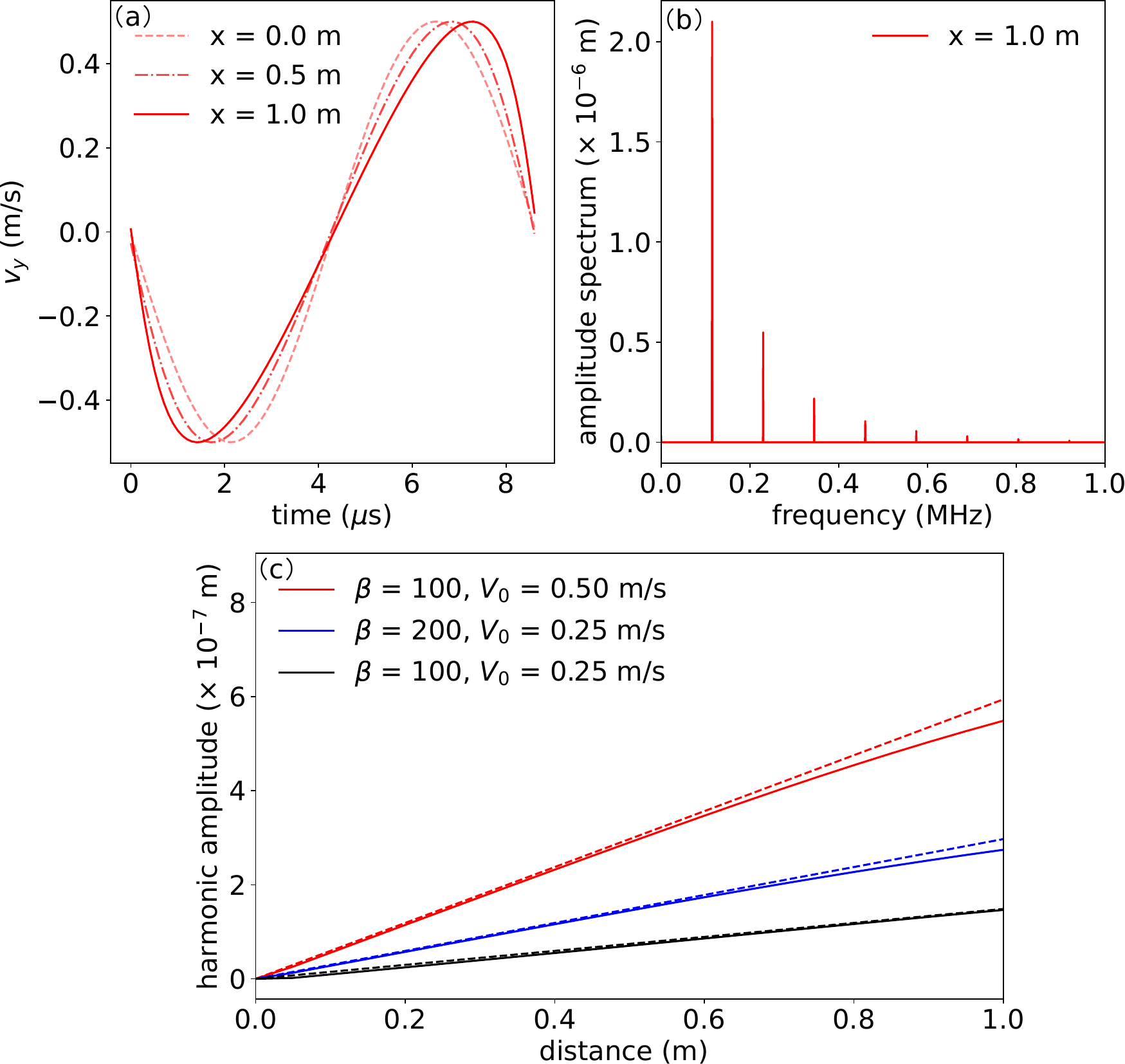} 
    \caption{\small Comparison between numerical and analytical asymptotic solutions for wave propagation from a single-frequency source. (a)  Recorded time series of $v_y$ at the source (dashed red curve) and at distances of 0.5 m (dash-dotted red curve) and 1.0 m (solid red curve) from the source. (b) The frequency amplitude spectrum of the time series of $v_y$ at 1.0 m from the source shows the generation of high-order harmonics, which are multiples of the fundamental frequency. (c) Comparison between the analytical asymptotic solutions (dashed curve) and the numerical (solid curves) solutions. We show three sets of parameters, with variations in the nonlinear modulus $\beta$ and the amplitude of the source $V_0$. We note that the analytical asymptotic solutions are known to be only valid at short distances from the source.}
    \label{fig:harmonics}
\end{figure}

The single-frequency waveform is modulated by the nonlinear parameter $\beta$ during propagation. Fig. \ref{fig:harmonics}a shows the modeled time series at distances of 0.0, 0.5, and 1.0 m from the source.
While the peak amplitude and period remain unchanged,
the shape of the waveform changes within one period due to the high-order harmonic generation. 

We show the generated harmonics 1.0 m away from the source in Fig. \ref{fig:harmonics}b. \citet{mccall1994theoretical} derived an asymptotic solution for the amplitude of the second-order harmonics at small distances away from the source. This analytical asymptotic solution is no longer valid at larger distances. As shown in Fig. \ref{fig:harmonics}c, these analytical solutions (dashed curves) serve as exact asymptotes to the numerical solutions (solid curves) at small distances. We present results for three sets of parameters, demonstrating the robustness of the match between the analytical asymptotic and our numerical solutions.

\newpage
\subsection{3D stress- and damage-induced anisotropy}
\label{subsec:veri_anisotropy}

Rocks exhibit various types and levels of anisotropy \citep{nur1969stress,nur1971effects,browning2017acoustic}. This anisotropy arises from various internal flaws, such as cracks, joints, and fabric development due to differential stress and strain during tectonic processes \citep{panteleev2024azimuthal}. The anisotropy of seismic wave propagation in such rocks can depend on the stress state and accumulated damage, a phenomenon referred to as stress- and damage-induced anisotropy. 
This dependence leads to nonlinear stress-strain relationships, which are important for capturing path and site effects in earthquake simulations. Accurately resolving these effects is essential to advance numerical simulations of ground motions. 

Both Murnaghan nonlinear elasticity and CDM describe stress-induced anisotropy \citep{johnson1993nonlinear,hamiel2009brittle}. However, while Murnaghan nonlinear elasticity may require unrealistically high values for $l_0$, $m_0$, and $n_0$ in Eq. (\ref{ivm-model}), CDM provides a physical framework that can describe stress- and damage-induced anisotropy and has been experimentally validated \citep{hamiel2009brittle}.
Here, we demonstrate that our proposed generic algorithm is suitable for implementing CDM by verifying its ability to resolve stress- and damage-induced anisotropy in 3D. We compare the numerical results with the analytical solutions derived by \citet{hamiel2009brittle}. 

We set up several plane-wave initial value problems to investigate how the P, S, and qS wave speeds depend on the orientation of the initial stress with respect to the normal vector of the initial wavefront and the damage level $\alpha$. The qS wave speed is the additional wave speed resulting from anisotropy \citep{harris2009scec}. Without loss of generality, we fix the normal vector of the wavefront to (1,0,0) and vary only the initial stress field and $\alpha$. 
Since CDM represents the seismic wave field using the total strain tensor
$\underset{=}{\varepsilon} = \underset{=}{\varepsilon}^{\text{pre}} + \underset{=}{\varepsilon}^{\text{dyn}}$, we pragmatically apply initial stress by prescribing initial strain values. 

The initial strain field consists of two parts: (i) a uniform strain field $\underset{=}{\varepsilon}^{\text{pre}}$, that represents the stress (strain) state of the rocks before dynamic perturbations from seismic waves; and (ii) the perturbation field $u^{\text{dyn}}_i = (\varepsilon_{xx}^{\text{dyn}},\varepsilon_{yy}^{\text{dyn}},\varepsilon_{zz}^{\text{dyn}},\varepsilon_{xy}^{\text{dyn}},\varepsilon_{yz}^{\text{dyn}},\varepsilon_{zx}^{\text{dyn}},v_x,v_y,v_z, \alpha)^T$, substituted into Eq. (\ref{nonlinear hyper pdes}). The expression for $u^{\text{dyn}}_i$ depends on the wave type and is given as 

\begin{equation}
    \begin{cases}
      u^{\text{dyn}}_i 
      = A_0 r^1_i \sin{(2 \pi k x)} &\text{, for P wave}\\
      u^{\text{dyn}}_i 
      = A_0 r^2_i \sin{(2 \pi k x)} &\text{, for S or qS wave} \\
      u^{\text{dyn}}_i 
      = A_0 r^3_i \sin{(2 \pi k x)} &\text{, for S or qS wave}
    \end{cases},
    \label{ani perturbation}
\end{equation}
where the three vectors $r_i^1$, $r_i^2$ and $r_i^3$ are defined in Eq. (\ref{q_b}). The classification of $r_i^2$ or $r_i^3$ is either S or qS waves depending on the orientation of the uniform strain field $\underset{=}{\varepsilon}^{\text{pre}}$.

We list the material properties of the CDM model and the initial values of the PDEs in Table \ref{tab:cdb_parameters}. The corresponding mathematical formulation is provided in Eq. (\ref{cdb-model}). We adopt the same cubic geometry as in Section \ref{subsec:veri_frequency}.

\begin{table}[hpt]
\centering
\caption{\small Summary of the perturbation field and the model parameters of the continuum damage model.}
\begin{tabular}{ M{2cm} P{2cm} P{2cm} P{1cm} P{2cm} P{2cm} P{1cm}}
\hline
 &Parameters &Values &Units &Parameters &Values &Units\\
\hline
\multirow{1}{*}{perturbations} 
&$A_0$ &2.5 $\times$ 10$^{-6}$ &1   &$k$ &20 &m$^{-1}$\\
\hline
\multirow{3}{*}{model para.} 
&$\lambda_0$ &32 &GPa   &$\gamma_r$ &37 &GPa\\
&$\mu_0$ &32 &GPa  &$\xi_0$  &-0.75 &1\\
&$\rho$ &2760 &kg/m$^3$ &$C_{d}$ &0.0 &(Pa$\cdot$s)$^{-1}$\\
\hline
\end{tabular}
\label{tab:cdb_parameters}
\end{table}

We set the initial damage variable $\alpha$ to 0.5. We define $\underset{=}{\varepsilon}^{\text{pre}}$ in its principal coordinate system as $(\varepsilon^{\text{pre}}_{xx}, \varepsilon^{\text{pre}}_{yy}, \varepsilon^{\text{pre}}_{zz}, \varepsilon^{\text{pre}}_{xy}, \varepsilon^{\text{pre}}_{yz}, \varepsilon^{\text{pre}}_{zx})^T = (1\times 10^{-3}, 0, 0, 0, 0, 0)^T$. Following \citet{hamiel2009brittle}, we initially align the global coordinate system in the numerical simulation with the principal coordinate system of $\underset{=}{\varepsilon}^{\text{pre}}$. We then rotate $\underset{=}{\varepsilon}^{\text{pre}}$ counterclockwise around the $z$-axis by an angle $\phi^{ani}$, which ranges from 0 to 180 degrees. 

\begin{figure}[hptb]
    \centering
    \includegraphics[width=1\columnwidth]{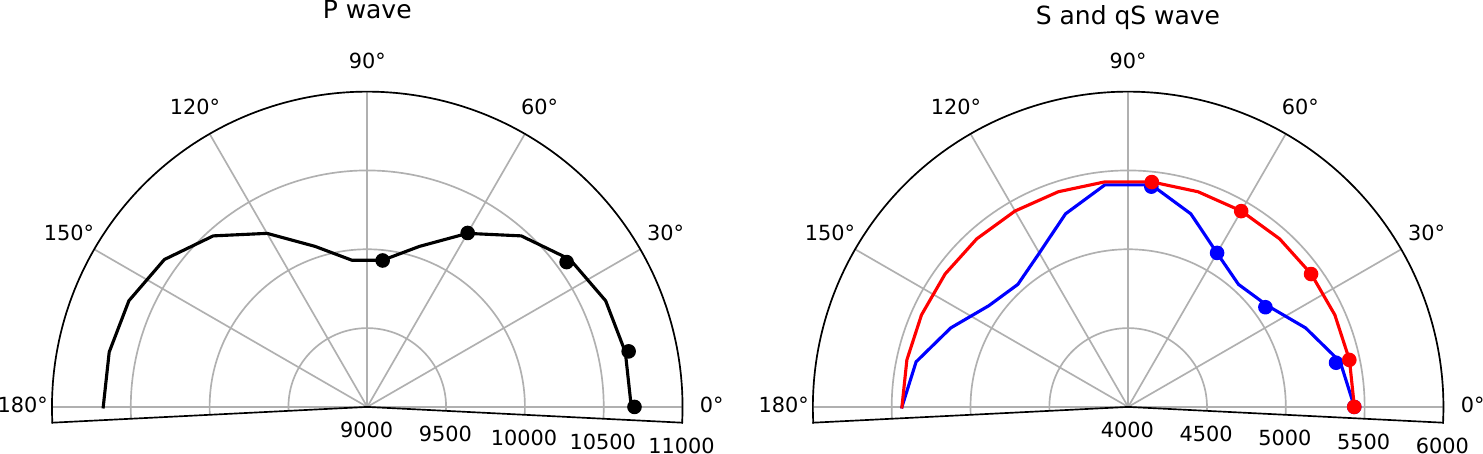}
    \caption{\small Comparison between analytical and numerical wave speeds of different phases for damage- and stress-induced anisotropy. (a) P-wave speed comparison, where black dots represent numerical simulation results and the black curve corresponds to the analytical solution. (b) S-wave (red curve and dots) and qS-wave (blue curve and dots) comparisons, showing numerical results alongside analytical predictions.}
    \label{fig:ani_angle}
\end{figure}

Figs. \ref{fig:ani_angle}a and \ref{fig:ani_angle}b compare analytical and numerical solutions for P waves and for S and qS waves, respectively. 

\section{Modeling co-seismic wave speed changes during the 2015 Gorkha earthquake}
\label{sec:kanthmandu_sim}

We apply our verified numerical framework to model co-seismic wave speed changes during the April 25, 2015, M$_w$ 7.8 Gorkha earthquake in the Kathmandu Valley.
We set up a geometrically complex 3D simulation of nonlinear seismic wave propagation from a finite source model of the 2015 M$_w$ 7.8 Gorkha earthquake.
Our setup captures key features relevant for modeling earthquake-related ground motions: a geometrically complex low-velocity sedimentary basin, layered subsurface geometry that represents different geological units, and a finite source model accounting for the directivity effect of a large earthquake.

\subsection{Numerical setup, nonlinear parameters and source model}

\begin{figure}[hptb]
    \centering
    \includegraphics[width=1\columnwidth]{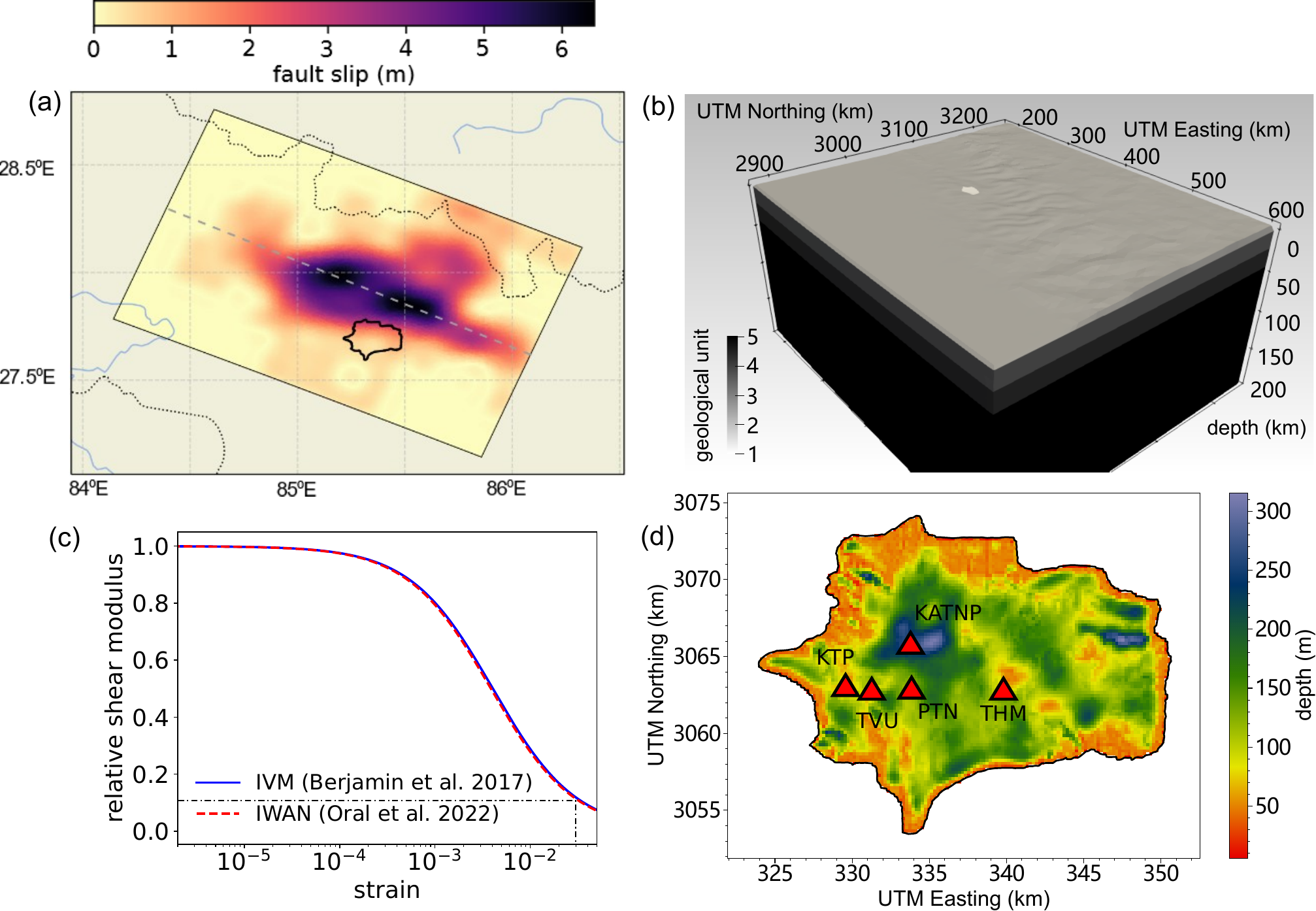}
    \caption{\small Model setup for the non-linear kinematic simulation of the 2015, $M_W$ 7.8 Gorkha earthquake. (a) Fault slip distribution interpolated from \citet{wei20182015}'s kinematic source model. The dashed gray line indicates the 12-km depth slice shown in Fig. \ref{fig:coseismic_changes}a. (b) Computational domain, consisting of five geological units. We incorporate topography, as well as the bathymetry of the sedimentary basin (white region at the upper boundary of the domain). 
    (c) Shear modulus reduction with strain amplitude of the IVM model (blue curve) within the basin that has been parameterized to match the IWAN model \citep[dashed red curve,][]{iwan1967class}. 
    (d) Map view of sedimentary basin depth variation, with five strong motion stations \citep{takai2016strong} marked by red triangles.}
    \label{fig:kathmandu_setup}
\end{figure}

As shown in Fig. \ref{fig:kathmandu_setup}b, the 3D computational domain has a size of 440×380×200~km$^3$. 
The velocity model includes five geological units (Table \ref{tab:kathmandu_parameters}). The first unit accounts for the surface topography and bathymetry of the shallow sediments within the Kathmandu basin with a low S-wave velocity of 200~m/s \citep{bohara2015analysis}. The second unit captures the strong topographical variation outside of the sedimentary basin within the Kathmandu Valley. We sample the surface topography with a resolution of 5~km. Units 3 through 5 are derived from a regional 1D velocity model \citep{mcnamara2017source}. 

We will compare the effects of three inelastic rheologies and elastic behavior using otherwise the same model setup: (i) visco-elastic, (ii) elasto-plastic, and (iii) internal variable model (IVM). In the visco-elastic case, we adopt the Zener model \citep{carcione1988wave} to describe viscous attenuation in SeisSol \citep{uphoff2016generating,uphoff_2024_14051105}. We list the visco-elastic quality factors for the P-wave ($Q_P$) and the S-wave ($Q_S$) inside each layer in Table \ref{tab:kathmandu_parameters}. The effective quality factors approximate the target quality factors well within the frequency range of 0.03 to 3 Hz. They increase asymptotically to infinity outside this frequency range, yielding close to linear elastic behavior. We set the quality factors as $Q_P$ = 0.1$V_S$ and $Q_S$ = 0.05$V_S$ for $V_S$ measured in m/s following \citet{olsen2003estimation}. In the elasto-plastic setup, the inelastic behavior is only effective inside the sedimentary basin (unit 1). We adopt the Drucker-Prager plasticity \citep{wollherr2018off} and provide the material parameters in the footnote of Table \ref{tab:kathmandu_parameters}.


We employ the IVM \citep{berjamin2017nonlinear} to investigate nonlinear co-seismic wave speed changes outside the fault core and extending over 100 kilometers from the fault.
The model has been validated in \citet{niu2024modeling} against two sets of laboratory experiments, which demonstrates its ability to quantify nonlinear co-seismic wave speed changes in granite samples \citep{manogharan2022experimental} and sandstone samples \citep{feng2018short}. 
The mathematical description of IVM nonlinearity is summarized in Eq. (\ref{ivm-model}). We refer to \citet{berjamin2017nonlinear} and \citet{niu2024modeling} for more details. The chosen model parameters of the IVM within each region are given in Table \ref{tab:kathmandu_parameters}. The nonlinear parameters inside the sedimentary basin (unit 1) are calibrated to match the modulus reduction curve from a 2D analysis presented in \citet{oral2022kathmandu}, constrained by the shift in resonance frequencies observed during significant events with magnitudes exceeding $M_W$ 6.5 within the Kathmandu Valley \citep{rajaure2017characterizing}.
For the layered bedrocks (units 2 to 5), we constrain the nonlinear IVM parameters from experiments by \citet{manogharan2022experimental} investigating nonlinear co-seismic wave speed changes of Westerly granite samples. As discussed in \citet{niu2024modeling}, the parameter $\gamma_b$, which determines the amplitude of stationary wave speed reductions under dynamic perturbations, can be constrained from experiments. However, the time scale $\tau_b$, which governs how quickly rocks reach the stationary state, remains highly uncertain. Here, we assume $\tau_b$ = 10 s in units 1 to 5, which is consistent with the time scale at which the changes in wave speed stabilize, as observed in experiments on Westerly granite samples \citep{manogharan2021nonlinear}.

\begin{table}[hptb]
\centering
\caption{\small Material parameters for each geological unit of the computational domain.}
\begin{threeparttable}
\begin{tabular}{ M{1cm} P{1.8cm} P{1.2cm} P{1.2cm} P{1.2cm} P{1.2cm} P{1.2cm} P{1cm} P{0.5cm}}
\hline
region &depth &$c_p$ &$c_s$ &$\rho$ &$Q_p$ &$Q_s$ &$\gamma_b$ &$\tau_b$\\
\hline
unit &km &m/s &m/s &kg/m$^3$ &1 &1 &kPa &s\\
\hline
1\tnote{*} &variable &300 &200 &1400 &20 &10 &0.5 &10 \\

2 &variable - 3 &5500 &3250 &2700 &325 &162.5 &356 &10 \\

3 &3 - 23 &5502 &3600 &2700 &360 &180 &437 &10 \\

4 &23 - 45 &6100 &3600 &2900 &360 &180 &437 &10 \\

5 &45 - 200 &8100 &4500 &3300 &450 &225 &550 &10 \\
\hline
\end{tabular}
\begin{tablenotes}\footnotesize
\item[*] Plasticity is only effective inside the sedimentary basin in the elasto-plastic simulation. The yielding strength is 224 kPa, with an internal friction angle of 26 degrees and a visco-plastic relaxation time $T_v$ of 0.05 s \citep{wollherr2018off}.
\end{tablenotes}
\end{threeparttable}
\label{tab:kathmandu_parameters}
\end{table}

We employ a polynomial degree of five and SeisSol's velocity-aware meshing capabilities to adapt the element size $h$ for each of the five geological units, ensuring at least three elements per S-wave wavelength of a maximum target frequency. In this way, our simulations resolve up to 0.5 Hz of the seismic wavefield everywhere in the domain, including in the complex geometry, low-velocity basin. 
We refine this mesh around the finite fault plane, which is embedded in units 1 to 3, to $h$ = 800 m for a higher resolution of the kinematic rupture evolution. As a result of the velocity-aware meshing, the sedimentary basin (unit 1) is resolved with a higher mesh resolution of $h$ $\approx$ 133 m. In units 2 and 3, mesh resolution gradually decreases, and $h$ increases from 800~m near the finite fault plane to $\approx$2000~m away from the source region.

In this example, we implement the finite source model of \citet{wei20182015} on a meshed finite fault plane to represent the $M_W$ 7.8 Gorkha earthquake.  
We do not model the spontaneous dynamic rupture process on the fault. The relatively coarse kinematic source model is interpolated using 2D polynomial functions of degree three over a 186 km $\times$ 121 km rectangular fault plane, which results in 22,506 square sub-faults of size 1 km $\times$ 1 km. We infer a variable slip rate on each of these sub-faults from the finite source model. Next, we interpolate the imposed slip rates onto SeisSol's triangular fault mesh as an internal boundary condition. This implementation is based on the approach by \citet{tinti2005earthquake,causse2014variability}. We use a Gaussian source time function to describe the slip rate function on each fault element \citep{bouchon1997state}.

\subsection{Large-scale nonlinear co-seismic wave speed changes}

\begin{figure}[hptb]
    \centering
    \includegraphics[width=1\columnwidth]{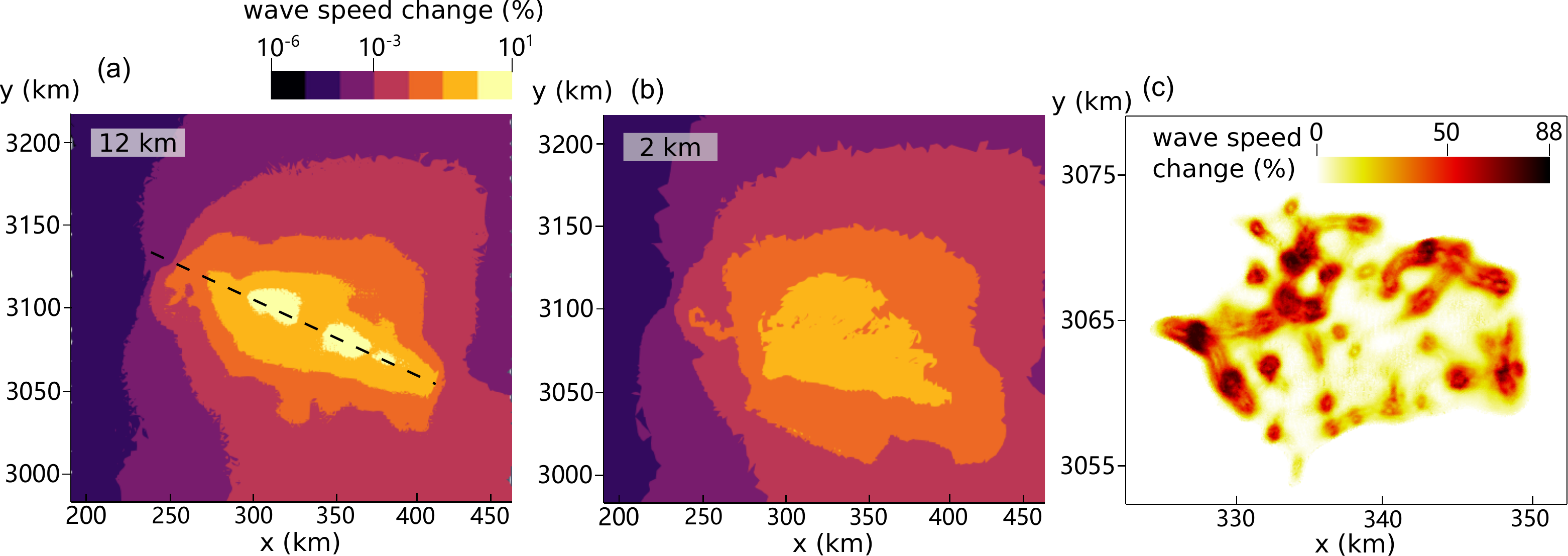}
    \caption{\small Map views of co-seismic wave speed changes and fault slip distribution.  (a) Co-seismic wave speed changes at 12 km depth, illustrating spatial variations in velocity reduction. The dashed black line marks the fault plane location at this depth. 
    (b) Co-seismic wave speed changes at 2 km depth, highlighting near-surface variations in wave speed reduction.
    (c) Co-seismic wave speed changes within the sedimentary basin, showing localized effects of nonlinearity in low-modulus materials. }
    \label{fig:coseismic_changes}
\end{figure}

Our nonlinear simulations reveal a significant reduction of co-seismic wave speed changes following the Gorkha earthquake across a vast region (Fig. \ref{fig:coseismic_changes}). Fig. \ref{fig:coseismic_changes}a shows wave speed changes 80~s after the rupture onset at 12~km depth. Nonlinear co-seismic wave speed reductions near the source range between 1\% and 10\% and are particularly pronounced close to the fault plane.
For example, in the 12-km depth slice shown in Fig. \ref{fig:coseismic_changes}a), the dashed black line marks the fault plane, which hosts a high slip at this depth.

The spatial distribution of the near-fault wave speed changes correlates with the fault slip distribution (Fig. \ref{fig:kathmandu_setup}a), with larger reductions in areas of large fault slip. Within the range of 70 km from the fault intersection, the wave speed reductions all exceed 0.01\%. This level of damage is still measurable with coda-wave- or ambient-noise-based interferometry \citep[e.g.,][]{brenguier2014mapping,gassenmeier2016field,lu2022regional}.

We show simulated co-seismic wave speed changes at 2 km depth in Fig. \ref{fig:coseismic_changes}b, which are lower compared with the changes at 12 km depth in Fig. \ref{fig:coseismic_changes}a.
However, the affected region is larger. At 2~km depth, wave speed reductions exceed 0.01\% within a 100~km radius. 

Within the sedimentary basin, nonlinear co-seismic wave speed changes are much larger (Fig. \ref{fig:coseismic_changes}c), and peak changes reach 88\%, corresponding to local peak strains up to 3$\times$10$^{-2}$ as can be seen in the shear modulus reduction curve (Fig. \ref{fig:kathmandu_setup}c). 
The spatial distribution of these changes correlates with the depth variations of the sedimentary basin (Fig. \ref{fig:kathmandu_setup} d), with greater reductions in wave speed located in regions with larger basin depths.
These findings align with field observations of nonlinear site effects, which report significant wave speed reductions in soft sediments during strong shaking \citep{bonilla2011nonlinear}.
We will further compare the wave speed changes modeled here with observations in Section \ref{sec:discussion}.

\subsection{Nonlinear site effects and sedimentary basin effects}

\begin{figure}[hptb]
    \centering
    \includegraphics[width=0.9\columnwidth]{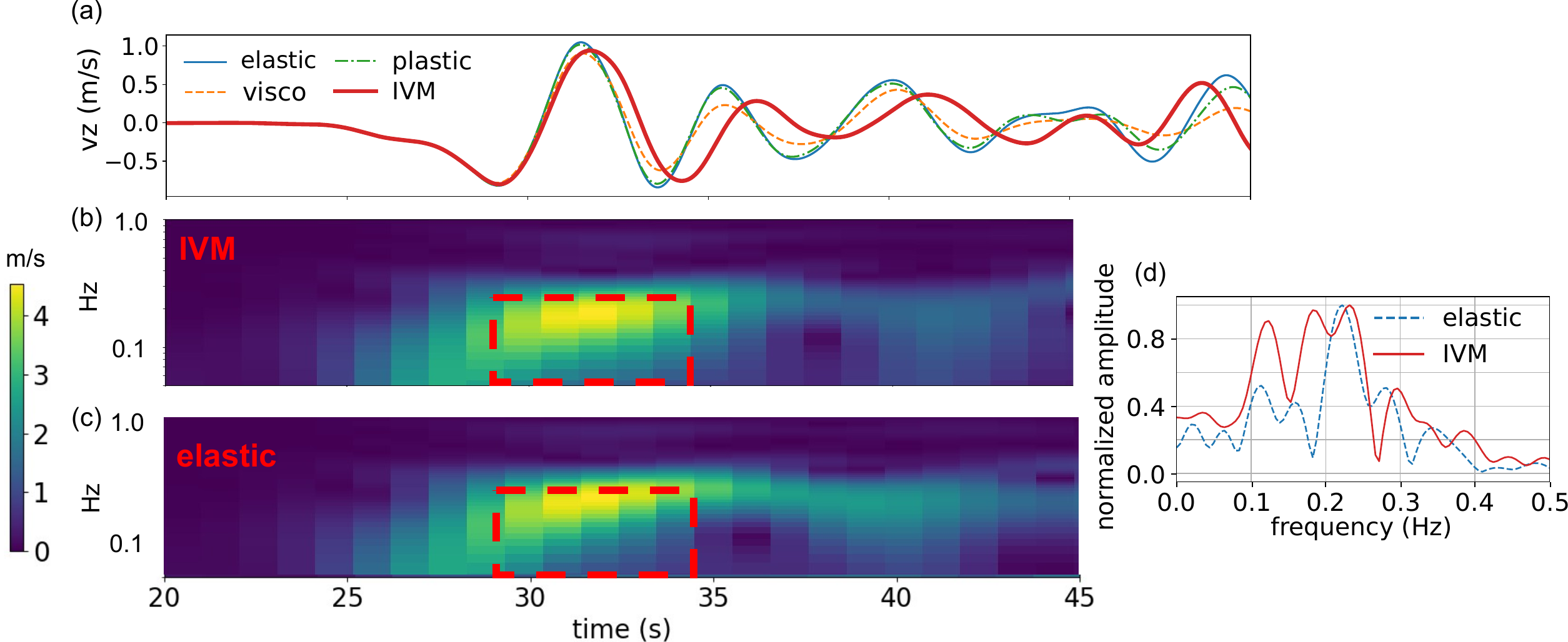}
    \caption{\small Time series and frequency analysis at station KTP. (a) Time series recorded at station KTP (marked in Fig. \ref{fig:kathmandu_setup}b) for different rheological models: elastic (solid blue curve), elasto-plastic (dash-dotted green curve), visco-elastic (dashed orange curve), and the IVM (solid red curve). 
    (b) and (c) are spectrograms of the IVM and elastic cases, respectively, showing the frequency content of the recorded waveforms. The dashed red rectangles highlight the amplification of lower-frequency components in the IVM simulation. 
    (d) Normalized frequency spectra of the time series recorded between 20 s and 50 s, comparing elastic (dashed blue curve) and IVM (solid red curve) models, illustrating the enhanced low-frequency content in the IVM simulation. In Fig. \ref{fig:kathmandu_freq_more}, we show the frequency spectra of time series recorded at four other stations marked in Fig. \ref{fig:kathmandu_setup}d.}
    \label{fig:kathmandu_time_series}
\end{figure}

In conjunction with co-seismic wave speed changes, we observe clear effects of the nonlinear rheology on ground motions. 
Such effects are exemplified in synthetic seismograms comparing linear elastic, visco-elastic, perfect elasto-plastic, and nonlinear damage model simulations (Fig. \ref{fig:kathmandu_time_series}a) at station KTP (Fig. \ref{fig:kathmandu_setup}d).
Compared to the linear elastic case, all three other models show different levels of ground motion damping at station KTP. 
The nonlinear damage model exhibits the strongest wave attenuation due to progressive modulus degradation, the accumulation of damage leading to the reduction of moduli. 

Our simulations suggest that co-seismic degradation of rock moduli may be an important mechanism contributing to the observed low-frequency amplification in soft sediments \citep{bonilla2011nonlinear}. We capture this effect in the spectrograms of nonlinear damage vs. linear elastic models (Figs. \ref{fig:kathmandu_time_series}b, c).
In the amplitude-frequency spectra of the modeled ground motion recorded between 20~s and 50~s after rupture onset(Fig. \ref{fig:kathmandu_time_series}d), we observe a systematic enhancement of low-frequency components (0.1--0.2 Hz). In our simulation, this low-frequency amplification is not unique to station KTP. As shown in Fig. \ref{fig:kathmandu_freq_more}, low-frequency amplification is a general feature of the modeled
ground motions at stations with high PGV values. High PGVs are correlated with significant ground deformation, leading to strong moduli reduction, consistent with the IVM shear modulus reduction curve (Fig. \ref{fig:kathmandu_setup}c). Such low-frequency amplification is expected during wave propagation through materials with co-propagating wave speed reduction. For example, a laboratory acoustic experiment on rock samples illustrates this phenomenon \citep{remillieux2017propagation}, where wave speed reduction delays the arrival time of later phases, elongating the period and consequently shifting the energy to a lower frequency.

\subsection{Nonlinear rheology and ground motions ($<$0.5 Hz)}

\begin{figure}[hptb]
    \centering
    \includegraphics[width=0.9\columnwidth]{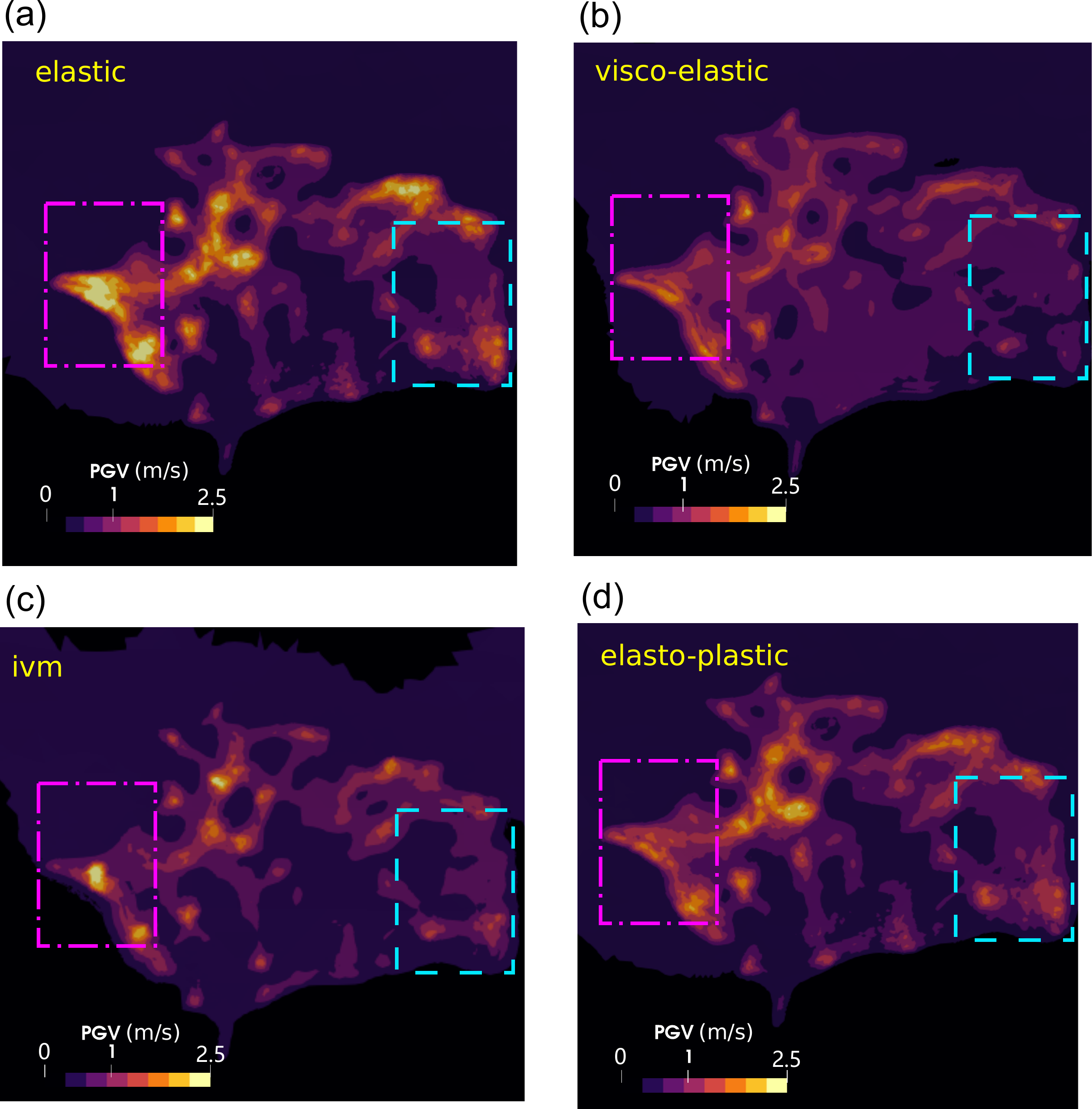}
    \caption{\small Maps of peak ground velocity (PGV) for different rheologies: (a) elastic, (b) visco-elastic, (c) IVM and (d) elasto-plastic. The dashed blue rectangles highlight the region where the elasto-plastic model exhibits minimal attenuation, while the dash-dotted pink rectangles indicate areas where attenuation is more pronounced.}
    \label{fig:kathmandu_pgvs}
\end{figure}

We compare modeled shake maps of peak ground velocity (PGV) across models with varying rheologies in Fig. \ref{fig:kathmandu_pgvs}. 
Linear elastic simulations show a strong correlation between the PGV in Fig. \ref{fig:kathmandu_pgvs}a and the depth of the sedimentary basin in Fig. \ref{fig:kathmandu_setup}d. 
Visco-elastic and elasto-plastic models reduce PGVs inside the Kathmandu basin, consistent with previous regional-scale studies \citep{narayan2014three,taborda2015physics,esmaeilzadeh20193d}.  Extending Southern California ShakeOut simulations to include IWAN plasticity also led to a reduction in ground motion amplitudes \citep[e.g.,][]{roten2023implementation}.

The nonlinear damage model attenuates PGVs across both high- and low-shaking intensity regions, unlike the elasto-plastic model, which primarily reduces high PGVs (Fig. \ref{fig:kathmandu_pgvs}b). 
The elasto-plastic model attenuates regions of high PGVs, such as in the pink dash-dotted rectangles in Fig. \ref{fig:kathmandu_pgvs}b. However, elasto-plastic effects are negligible in regions with relatively low PGVs, such as those marked with blue dashed rectangles in Fig. \ref{fig:kathmandu_pgvs}b, which is expected from previous theoretical work and numerical simulations \citep[e.g.,][]{roten2014expected,kojima2016closed,seylabi2021deterministic}. The plastic yielding surface is only reached when stress reaches a certain threshold. Below this threshold, the mechanical behavior of the material is the same as that of the linear elastic model.
In contrast, the nonlinear damage model continuously degrades moduli with increasing strain amplitude (Fig. \ref{fig:kathmandu_setup}c). 

\section{Discussion}
\label{sec:discussion}

\subsection{Accuracy and performance of the nonlinear implementation}
\label{subsec:discussion_accuracy_and_efficiency}

In Section \ref{sec:kanthmandu_sim}, we applied the proposed algorithm to model regional-scale nonlinear co-seismic wave speed changes in 3D. Nonlinear seismic wave propagation simulations are computationally demanding, necessitating efficient algorithms and optimized implementations for execution on large-scale high-performance computing (HPC) systems \citep[e.g.,][]{reinarz2020exahype,roten2023implementation}. To illustrate the efficiency of our nonlinear PDE solver, we analyze its convergence rate with reduced element size $h$ in Section \ref{subsec:veri_riemann}. We also analyze $p$ convergence in Fig. \ref{fig:riemann_problem}, where the $L_2$ errors in numerical solutions decrease with element shape functions of higher polynomial degree $p$.
 
Fig. \ref{fig:riemann_problem} shows a first-order convergence rate for simulations using basis functions of polynomial degrees 1 to 5. This low order of convergence results from the linearized Cauchy-Kovalevskaya procedure used in the prediction step, c.f. Eq. (\ref{2nd derivative}). The prediction step approximates the time-dependent coefficients $U_{lp}(t)$ within a single time step using a Taylor series expansion \citep{toro2001towards}. In this step, to compute high-order time derivatives, we linearize the nonlinear hyperbolic equations in Eq. (\ref{2nd derivative}) and apply the Cauchy–Kovalevskaya procedure to the linearized system, as detailed in \citet{dumbser2006arbitrary}. This linearization ensures algorithmic generality across various nonlinear rock models but limits the accuracy of $U_{lp}(t)$ at higher orders, thus constraining the overall convergence rate.

A low-order convergence rate observed at solution discontinuities, such as shock waves, is consistent with Godunov’s theorem \citep{godunov1959finite}. This theorem establishes that high-order linear solvers have non-monotonic behavior near steep solution gradients. In addition, spectral convergence properties might be reduced to low-order accuracy due to the manifestation of the well-known Gibbs phenomena in the vicinity of strong discontinuities \citep[e.g.,][Chapter 5.6]{hesthaven2007nodal}.  Local low-order convergence is also evident in SeisSol's dynamic rupture implementation \citep[Sec. 6.3][]{wollherr2018off}. 

A potentially promising extension of our work is the incorporation of a discrete Picard iteration scheme \citep{lindelof1894application,youssef2007picard,dumbser2008unified,gassner2011explicit,reinarz2020exahype}. The Picard iteration can substitute our linearized Cauchy-Kovalevskaya procedure in the prediction step to estimate $\mathcal{D}_{lp}^i$ in Eq. (\ref{cdb-model}). This approach has been shown to help preserve high-order convergence up to 7 in ADER-DG solvers \citep{dumbser2008unified}.

We analyze the performance of our SeisSol implementation on the supercomputer Frontera at TACC \citep{stanzione_frontera_2020}.
Additionally, we suggest potential improvements to enhance the current algorithm, including future large-scale hardware architectures. 

\begin{figure}[hptb]
    \centering
    \includegraphics[width=1.0\columnwidth]{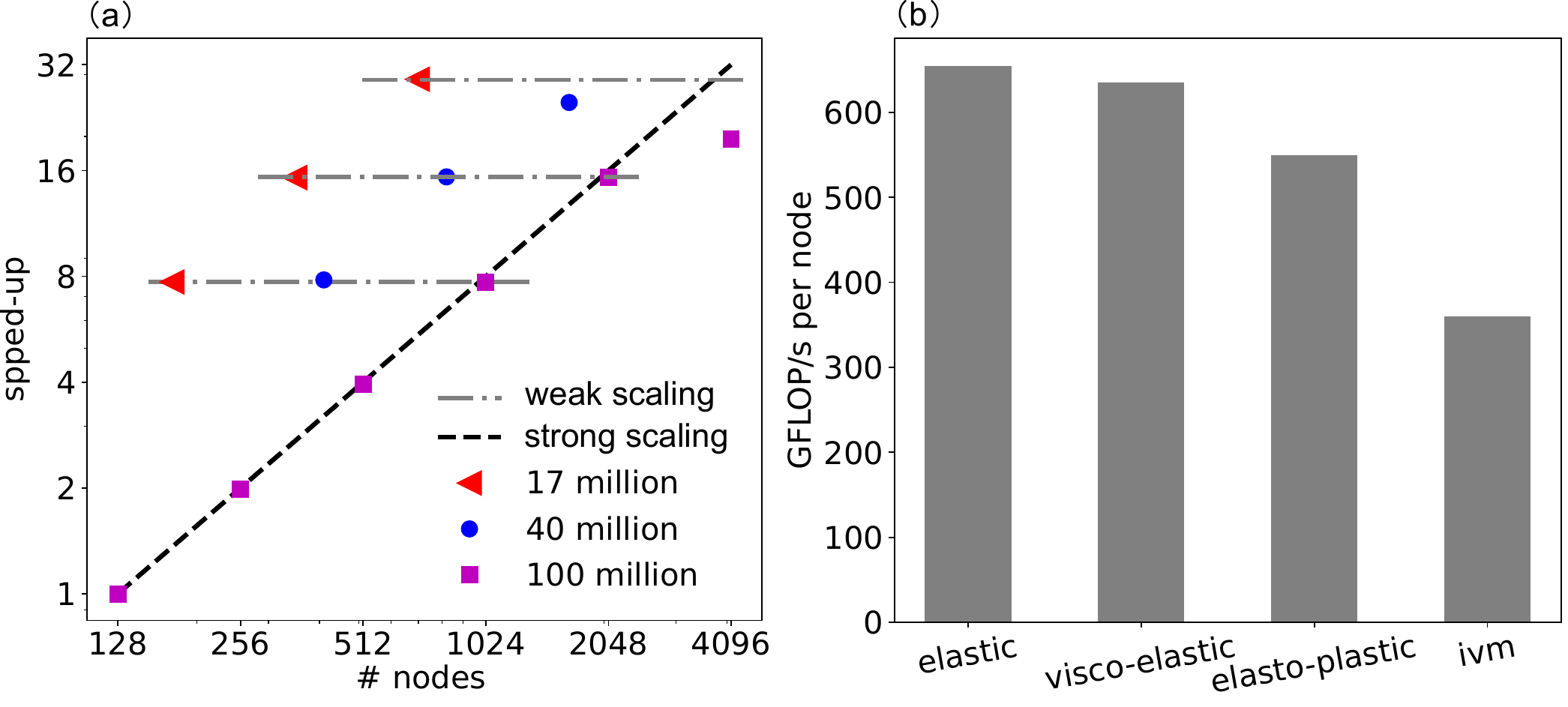}
    \caption{\small Scalability and performance. 
    (a) Speed-up of simulations as a function of the number of compute nodes, scaling up to 4096 nodes on Frontera \citep{stanzione_frontera_2020}. The dashed black curve represents the ideal strong-scaling regime, where doubling the number of nodes halves the time to solution. The dash-dotted gray curves illustrate the ideal weak-scaling regime, where proportionally increasing the number of nodes with the number of mesh elements results in the same speed-up. Different mesh sizes are represented by red triangles (17 million elements), blue circles (40 million elements), and purple rectangles (100 million elements). Both axes use a logarithmic scale.
    (b) Hardware performance analysis during simulations of the 2015 $M_W$ 7.8 Ghorka earthquake (Section \ref{sec:kanthmandu_sim}) for different rock models, shown as a bar plot. The mesh used here contains $\approx$2.3 million elements, and the simulation ran on 32 nodes of SuperMUC-NG (Phase 1).}
    \label{fig:performance}
\end{figure}

We evaluate the scalability and speed-up of the nonlinear SeisSol implementation using the 2015 Kathmandu earthquake model shown in Fig. \ref{fig:kathmandu_setup}b. We here discretize the simulation domain with three different meshes containing approximately 17, 40, and 100 million elements, respectively. In the discontinuous Galerkin (DG) method, the degrees of freedom (DOFs) are directly proportional to the number of tetrahedral elements. We use a polynomial degree $p=3$ (Eq. \ref{modal dg space}) for performance analysis, resulting in 200 DOFs per element.

The scaling tests consist of simulations using all three meshes and various numbers of compute nodes, running for 3 s of physical simulation time with the same time step size. SeisSol employs a hybrid MPI-OpenMP parallelization scheme, utilizing MPI for inter-node communication and OpenMP for multi-threaded parallelization within each node \citep{uphoff_extreme_2017}.

We evaluate the performance in terms of speed-up, which is defined as $t_s / t_0$ with $t_s$ being the time to solution for a given combination of mesh size and number of compute nodes, $t_0$ is the time to solution of the baseline simulation which uses a 100-million-element mesh on 128 nodes. Fig. \ref{fig:performance} illustrates the scalability on the Frontera supercomputer at TACC \citep{stanzione_frontera_2020}. Frontera employs Intel Xeon Platinum 8280 ("Cascade Lake") processors, each offering 56 cores per node and operating at 2.7 GHz. The total number of available compute nodes is 8,368.

We analyze how speed-up depends on mesh sizes and the number of compute nodes in Fig. \ref{fig:performance}a. To facilitate direct comparison across different mesh sizes for both strong and weak scaling, we normalize the speed-up by nodes per million elements in the following discussions.
The results indicate that for fewer than 20 nodes per million elements, strong scaling is nearly linear using the 100 million element mesh, meaning that speed-up increases almost proportionally with node count. 

To analyze weak scaling behavior, we compare different mesh sizes using the same number of nodes per million elements. The speed-up across the three different mesh sizes remains nearly identical as long as the number of nodes per million elements remains below 20. However, at 40 nodes per million elements, performance deviates significantly from ideal scaling in both strong and weak scaling tests. Performance degradation becomes more pronounced as the number of elements increases, corresponding to a larger number of compute nodes. One possible explanation is that the communication time between MPI ranks occupies a larger proportion of the overall computation time. Optimizing SeisSol’s performance at those higher node counts is beyond the scope of this study and requires further development efforts.

We compare the performance of our implementation using nonlinear space-time interpolation kernels with that of existing SeisSol models. Since our implementation in this work for nonlinear hyperbolic equations only supports a uniform time step size across the entire simulation domain (global time stepping, GTS), we constrain our comparison with the other existing models in SeisSol to the GTS scheme.
\citet{uphoff_extreme_2017} demonstrate the strong scaling behavior of SeisSol for dynamic rupture earthquake simulations using a linear elastic model. With a mesh containing approximately 51 million elements, the parallel efficiency remained $\sim$95\% on 512 nodes compared to a performance of $\sim$660 GFLOP/s on 16 nodes. The simulation on 512 nodes corresponds to $\sim$10 nodes per million elements, which is within the range of our scaling analysis in Fig. \ref{fig:performance}a.

In terms of strong scaling, our nonlinear implementation reaches a speed-up of $\sim$15.3 when increasing the number of nodes from 128 to 2048 for a mesh with $\sim$100 million elements. This result is comparable to the elastic model above, with a parallel efficiency of 95.7\% up to $\sim$20 nodes per million elements. However, when the number of nodes is further increased to 4,096, the parallel efficiency drops to 61.5\%,  indicating the need for further optimization of our current implementation for handling nonlinear wave propagation at extreme scales. For example, \citet{wolf2022efficient} recently optimized the implementation of computationally intensive poro-elastic rheologies in SeisSol, achieving performance degradation of less than 10\%, even at more than 40 nodes per million elements. 

The strong scaling behavior does not fully capture the absolute performance of the code in terms of floating point operations per second (FLOP/s). To provide a more precise assessment, we compare FLOP/s among simulations using the four material models described in Section \ref{sec:kanthmandu_sim}. For a 2.3 million element mesh, performance measurements are taken from results running on 16 nodes of SuperMUC-NG (Phase 1) with shape functions of polynomial degree 3. SuperMUC-NG employs Intel Xeon Platinum 8174 processors, each equipped with 48 cores per node, operating at 2.7 GHz. As shown in Fig. \ref{fig:performance}b, simulations with elastic, visco-elastic, and elasto-plastic materials achieve a node-average performance of 654 GFLOP/s, 636 GFLOP/s, and 550 GFLOP/s, respectively, using double-precision floating-point arithmetic. In contrast, the nonlinear implementation with IVM achieves 360 GFLOP/s, which represents a 45\% reduction in computational performance compared to the elastic model. 

The current implementation does not yet support local time stepping \citep[LTS,][]{breuer2016petascale,uphoff2020flexible}, which is crucial for efficiently handling non-uniform element sizes due to mesh refinement near faults, complex fault geometries, or highly-varying surface topography. Thus, on the same mesh, the time-to-solution for the nonlinear IVM implementation is approximately 5.56 times longer than the linear elastic material in our simulations presented in Section \ref{sec:kanthmandu_sim}. Therefore, future implementation of LTS for nonlinear models is a promising avenue for improving computational efficiency while maintaining accuracy.

\subsection{Linking co-seismic wave speed changes of rocks from laboratory measurements to regional scale field observations}
\label{subsec:discuss_coseismic_wave_speed_changes}

In this section, we discuss what the simulations of the 2015 Ghorka earthquake reveal about co-seismic wave speed changes in linking measurements of co-seismic wave speed changes from the laboratory with field-scale observations. Under well-controlled environments and boundary conditions in the laboratory, the dynamic responses of rocks to seismic wave fields can be better constrained. In this work, we employ an experimentally constrained continuum mechanics model, the IVM \citep{berjamin2017nonlinear,niu2024modeling}.
However, the amplitudes of the modeled regional wave speed changes may not be comparable to observations during the 2015 Ghorka earthquake.
In the following, we discuss reasons that may contribute to the amplitude difference between the simulated regional co-seismic wave speed changes and those in field observations.

\citet{lu2022regional} show that the average wave speed changes within a depth range from 0 to $\approx$3 km can exceed 1\% within 90 km from the fault. These changes are two orders of magnitude larger than our simulated wave speed changes at depths of 2 km within 100 km from the fault, which is likely due to large perturbations within soft sediments across the upper few hundred meters below the surface. Such significant perturbations inside the sediments are not reflected in our analysis of a depth slice at 2 km. Fig. \ref{fig:coseismic_changes}c shows that wave speed changes within the sedimentary basin reach 88\%. Similarly, using seismic observations from the KiK-net network, \citet{bonilla2019monitoring} observe wave speed reductions greater than 60\% in shallow soft sediments within 150 s after the occurrence of the 2011 $M_W$ 9.0 Tohoku-oki earthquake in Japan. These results suggest that incorporating the shallowest sedimentary layers may increase the average wave speed changes, potentially enabling a more quantitative comparison between numerical simulations and field observations. 

Although this study demonstrates how to adapt laboratory-derived nonlinear models to regional-scale numerical simulations of co-seismic wave speed changes, the nonlinear IVM material properties used in our simulations were not constrained with rock samples from the Kathmandu Valley. However, the spatial variation patterns of co-seismic wave speed changes modeled here may be transferable across similar lithologies. 
For example, our simulations reveal that the amplitude of co-seismic wave speed changes correlates strongly with fault slip close to the source (Figs. \ref{fig:coseismic_changes}a and \ref{fig:kathmandu_setup}a). At increasing distances from the fault, the dynamic strain amplitude is modulated by the layered Earth model, shown in Fig. \ref{fig:kathmandu_setup}. With slightly softer rocks (lower $c_s$ in Table \ref{tab:kathmandu_parameters}) at a depth of 2 km, the region where the changes in wave speed are greater than 0.01\% is broader than that at a depth of 12 km (Fig. \ref{fig:coseismic_changes}a and \ref{fig:coseismic_changes}b). This effect is particularly prominent within the sedimentary basin, where low-moduli unconsolidated materials experience greater strain amplification. We find that the basin depth distribution is an additional factor that adds to the spatial variability of changes in nonlinear wave speed. Our results (Fig. \ref{fig:coseismic_changes}c) indicate that larger sedimentary basin depths lead to greater co-seismic wave speed reductions. Other factors that might contribute to the variation, for example, the direction of incoming waves \citep{oral2022kathmandu}, require further investigation as a next step.

A limitation of our approach is that the nonlinear damage model (IVM) remains isotropic even as damage accumulates. However, material anisotropy may develop under high damage levels (Fig. \ref{fig:ani_angle}), further influencing directivity effects and path and site effects. Accounting for non-linear anisotropy will introduce additional challenges in accurately implementing free-surface boundary conditions. Although the method outlined in Section \ref{subsec:boundary_conditions} is suitable for isotropic models only, it can serve as a first-order approximation for damage- and stress-induced moduli changes at the free-surface boundary by only accounting for the induced changes in the effective Lam\'{e} parameters in Eq. (\ref{effective-cdb}).

\subsection{Incorporating background stress effects on co-seismic non-linear wave speed changes}

In Section \ref{sec:kanthmandu_sim}, we use the IVM with experimentally constrained parameters \citep{niu2024modeling} on Westerly granite to quantify the spatial distribution of co-seismic wave speed reductions following the 2015 $M_W$ 7.8 Ghorka earthquake. This model assumes a universal co-seismic wave speed reduction, irrespective of the initial stress state. Similar universal reductions in wave speed under dynamic perturbations have been observed in laboratory rock samples under unconfined stress conditions \cite{remillieux2017propagation,feng2018short} and under uniaxial compression of up to 20 MPa \citep{riviere2015set,manogharan2021nonlinear}.
However, \citet{manogharan2022experimental} show that the level of uniaxial compression exerts a second-order influence on the amplitude of co-seismic wave speed reductions, indicating that a more advanced model is needed to incorporate the dependence of wave speed changes on the background stress state.

The CDM \citep{lyakhovsky1997distributed}, described in Eq. (\ref{cdb-model}), explicitly accounts for the background stress state. In this model, the amplitude of damage accumulation depends on how close the current stress state is to a critical stress threshold, defined by $\xi_0$ in Eq. (\ref{cdb-model}). In Section \ref{subsec:veri_anisotropy}, we demonstrate that our proposed algorithm can quantify stress- and damage-induced anisotropy in wave propagation using CDM. However, applying CDM to co-seismic wave speed changes requires sufficient knowledge of the pre-existing background stress state.

Properly configuring the background stress state is especially important when modeling layered geological structures, particularly when accounting for spatially varying bathymetry in sedimentary basins (unit 1 in Fig. \ref{fig:kathmandu_setup}). Using CDM, the background stress state is imposed by specifying the initial strain tensor. To prevent spurious wave generation at the beginning of the simulation, it is necessary to ensure the stress continuity condition at layer boundaries. This is challenging when incorporating geometrically complex basin bathymetry, where the strain tensor must be reoriented according to the basin geometry. A potential solution to this challenge in future work may be first to solve the static strain field resulting from the overburden of rocks and soils. This balanced strain field may then be applied as the initial strain state for wave simulations, ensuring a physically consistent background stress distribution.

\section{Conclusions}
\label{sec:conclusion}
To develop a seismic wave propagation method capable of modeling observed co-seismic wave speed changes, we propose a generic numerical algorithm based on the discontinuous Galerkin (DG) method that can be applied to a wide range of nonlinear rock models.
We verify the numerical solutions obtained using our new approach implemented in the open-source software SeisSol against three sets of analytical solutions and confirm the convergence of the algorithm.
Using the Riemann problem setup, we demonstrate that the proposed method accurately resolves discontinuities in nonlinear hyperbolic equations. 
We find a 1st order convergence rate at solution discontinuities with basis functions of polynomial degrees 1 to 5. 
On the same mesh, using higher-degree basis functions leads to lower numerical errors.
We show that the method can accurately resolve the amplitude of high-frequency harmonics generated by wave propagation in the Murnaghan nonlinear elasticity model.
The proposed method can also properly quantify the stress- and damage-induced mechanical anisotropic behaviors of rocks.

We evaluate the parallel performance of our implementation on Frontera and find that both weak and strong scaling remain close to linear up to 20 nodes per million elements, allowing efficient simulations on meshes with up to 100 million elements and scalability up to 2048 nodes.
However, despite the good parallel scalability, node-level performance remains non-optimal, indicating the need for further optimizations to improve computational efficiency and reduce runtime for handling future nonlinear wave propagation simulations at extreme scales.

We apply our algorithm to regional-scale earthquake simulations, including non-linear wave propagation effects from source to site.
We use the experimentally constrained nonlinear model IVM to capture co-seismic wave speed changes during the 2015 M$_w$ 7.8 Gorkha earthquake in the Kathmandu Valley, incorporating a free surface with topography, a sedimentary basin with low wave speeds and complex bathymetry, a layered geological structure, and a finite source model that accounts for rupture directivity effects.
The simulation results show that co-seismic wave speed reductions depend on the fault slip distribution near the source and are modulated by basin depth tens of kilometers away from the fault.
Co-seismic wave speed changes also enhance low-frequency components in soft sedimentary layers, affecting ground motions.
This study demonstrates, using a physics-based framework to quantify nonlinear earthquake effects at a regional scale, the importance of damage-induced wave speed variations for seismic hazard assessment, ground motion predictions, and as an observable to better constrain earthquake physics and rock mechanics. 

\acknowledgments
The authors thank Dave A. May and Yehuda Ben-Zion for fruitful discussions. We are grateful to Ashim Rijal and Amrit Kaur for their contributions to the Kathmandu model setup.
This project has received support from the European Union's Horizon 2020 research and innovation programme under the Marie-Sklodowska-Curie grant agreement No. 955515 – SPIN ITN (www.spin-itn.eu) and the TEAR ERC Starting (grant No. 852992), from Horizon Europe (ChEESE-2P, grant no. 101093038; DT-GEO, grant no. 101058129; and Geo-INQUIRE, grant no. 101058518), the National Science Foundation and TACC's LCCF-CSA (grant numbers OAC-2139536, OAC-2311208, EAR-2225286, EAR-2121568), the National Aeronautics and Space Administration (NASA grant no. 80NSSC20K0495) and the Southern California Earthquake Center (SCEC projects \#21112, \#22135).
The authors acknowledge the Texas Advanced Computing Center (TACC) at The University of Texas at Austin for providing computational resources that have contributed to the research results reported in this paper (http://www.tacc.utexas.edu) and the Gauss Centre for Supercomputing e.V. (www.gauss-centre.eu) for providing computing time on the supercomputer SuperMUC-NG at the Leibniz Supercomputing Centre (www.lrz.de) in project pn49ha. Additional computing resources were provided by the Institute of Geophysics of LMU Munich \citep{oeser2006cluster}.

\section*{Data availability}

The source code of SeisSol with nonlinear IVM implementation is available as open-source software under \href{https://github.com/SeisSol/SeisSol/tree/damaged-material-nonlinear-drB}{https://github.com/SeisSol/SeisSol/tree/damaged-material-nonlinear-drB}. The model setup, simulation outputs, and post-processing scripts to reproduce all figures are available at a Zenodo \href{https://zenodo.org/records/14712559?preview=1&token=eyJhbGciOiJIUzUxMiJ9.eyJpZCI6Ijk5NzQ4NDA4LTMwYjctNGJiOC1hMDQ1LTE2YzIxYzRiNThmMiIsImRhdGEiOnt9LCJyYW5kb20iOiI1Yzc3OGM3NTY3MzlmMTk4MzY0Mzg0YTc5NjE0ZWM1YiJ9.Gfeu2H6k3bL52USggCSIVJfdpQ_5_Je5GW1JXpuWMAYAwYXccsyO3qJAMgbyh_qsGlWgAP0QRdMPs2e4EE7qTQ}{repository}.


\appendix

\section{DG algorithm for nonlinear wave equations}
\label{sec:app_dg_algorithm}

In this section, we provide the details on three components of the DG algorithm proposed in this work: prediction step, correction step, and boundary conditions.

\subsection{Prediction step: linearization and temporal approximation}
\label{subsubsec:prediction_step}

In the prediction step, we retain only the conservative term of Eq. (\ref{nonlinear hyper pdes}) assuming weak nonlinearity (${\partial \sigma_{ij}}/{\partial \varepsilon_{mn}}$ and ${\partial \sigma_{ij}}/{\partial \alpha}$ $\rightarrow$ constant) and employ a linearization procedure.
Our main motivation for this linearization in the prediction step is to maintain the HPC-optimized data structure of SeisSol \citep{uphoff_2024_14051105}. We will release this restriction in the subsequent correction step described later. This assumption preserves the convergence of the algorithm for nonlinear hyperbolic PDEs but can have an effect on the convergence rate, as we will discuss in Section \ref{subsec:veri_riemann}.

We write for the linearized prediction step: 
\begin{align}
    \dfrac{\partial u_p}{\partial t} 
    &= -\dfrac{\partial F^d_p}{\partial x_d} \nonumber \\
    &= -\dfrac{\partial F^d_p}{\partial u_q} 
        \dfrac{\partial u_q}{\partial x_d},
    \label{nonlinear-wave}
\end{align}
where $F_p^d=F_p^d(\underset{-}{u})$ is a nonlinear function of the conservative variables $u_p$, with $\dfrac{\partial F^d_p}{\partial u_q}$ corresponding to its Jacobian matrix. Taking a time derivative on both sides of Eq. (\ref{nonlinear-wave}), we approximate the second time derivative of $u_p$ as:

\begin{align}
    \dfrac{\partial^2 u_p}{\partial t^2} 
    &= -\dfrac{\partial}{\partial t} 
        \left( \dfrac{\partial F^d_p}{\partial u_q} 
        \dfrac{\partial u_q}{\partial x_d} \right) \nonumber \\
    &= - \dfrac{\partial}{\partial t} \left(\dfrac{\partial F^d_p}{\partial u_q}\right) \dfrac{\partial u_q}{\partial x_d}
       -\dfrac{\partial F^d_p}{\partial u_q} 
        \dfrac{\partial}{\partial x_d} \left(\dfrac{\partial u_q}{\partial t}\right)  \nonumber \\
    &\approx -\dfrac{\partial F^d_p}{\partial u_q} 
        \dfrac{\partial}{\partial x_d} \left(\dfrac{\partial u_q}{\partial t}\right).
    \label{2nd derivative}
\end{align}

This condition is satisfied if $\dfrac{\partial}{\partial t} \left(\dfrac{\partial F^d_p}{\partial u_q}\right) \dfrac{\partial u_q}{\partial x_d} \ll 
\dfrac{\partial F^d_p}{\partial u_q} 
\dfrac{\partial}{\partial x_d} \left(\dfrac{\partial u_q}{\partial t}\right)$, which requires $\dfrac{\partial F^d_p}{\partial u_q}$ to vary slowly in time compared to the temporal variation of $u_q$. 

From Eqs. (\ref{nonlinear damage models}) and (\ref{nonlinear hyper pdes}), $F^d_p$ incorporates the nonlinear stress-strain relationships. Consequently, $\dfrac{\partial F^d_p}{\partial u_q}$ changes gradually under weak nonlinearity. The weak nonlinearity makes Eq. (\ref{2nd derivative}) a more accurate approximation for the second-order time derivative of $u_p$. We reiterate that this assumption only pertains in the prediction step.

Following \citet{uphoff2020flexible}, the arbitrary order ($i$) derivative of $q_p$ in time ($\mathcal{D}_{lp}^i$) is computed as follows:

\begin{align}
    \mathcal{D}_{lp}^i \int_{\mathcal{T}_m} \phi_k \phi_l \text{d} V
      = - \int_{\mathcal{T}_m} \phi_k B_{pq}^d(\underset{-}{u}^{t_n}) \mathcal{D}_{lq}^{(i-1)} \dfrac{\partial \phi_l}{\partial x_d} \text{d} V,
    \label{CKIter}
\end{align}
where $\mathcal{D}_{lq}^i \phi_l = \dfrac{\partial^{i} u_q}{\partial t^i}$.

For linear wave equations, we derive $B_{pq}^d$ = $\dfrac{\partial F^d_p}{\partial u_q}$ as a cell-wise constant that keeps its value along the simulation \citep{uphoff2020flexible}. In our nonlinear case, we need to re-compute the cell-wise averaged $B_{pq}^d$ from $u_p^{t_n}$ at the beginning of each time step $t_n$, i.e. $B_{pq}^{d,t_n} = B_{pq}^d(\underset{-}{u}^{t_n}) = \int_{\mathcal{T}_m} B_{pq}^d(\underset{-}{u}^{t_n}) \text{d} V / V_{e}$ and $V_e$ is the volume of the tetrahedral element.

If we substitute $B_{pq}^{d,t_n}$ in Eq. (\ref{CKIter}), the integration in a reference cell $\mathcal{E}_3$, which is defined in a reference Cartesian coordinate system where the position vector of a point is $\xi_i$, will be
\begin{align}
    \mathcal{D}_{lp}^i |J| \int_{\mathcal{E}_3} \phi_k \phi_l \text{d} V
      = - |J| \Theta^{-1}_{ed} \mathcal{D}_{lp}^{(i-1)} B_{pq}^{d,t_n} \int_{\mathcal{E}_3} \phi_k \dfrac{\partial \phi_l}{\partial \xi_e} \text{d} V,
    \label{CKIter-ref}
\end{align}
where $\Theta^{-1}_{ed} = \partial \xi_e / \partial x_d$. We refer to Chapter 3.1 of \citet{uphoff2020flexible} for the detailed definition of the reference Cartesian coordinate system. Defining $M_{kl} = \int_{\mathcal{E}_3} \phi_k \phi_l \text{d} V$ and $K^e_{lk} = \int_{\mathcal{E}_3} \phi_k \dfrac{\partial \phi_l}{\partial \xi_e} \text{d} V$, we derive

\begin{align}
    \mathcal{D}_{lp}^i |J| M_{kl}
      = - |J| \Theta^{-1}_{ed} \mathcal{D}_{lq}^{(i-1)} B_{pq}^{d,t_n} K^e_{lk},
    \label{CKIter-ref-matrix}
\end{align}
which is directly comparable to Eq.(3.31) in \citet{uphoff2020flexible}.

If the nonlinear source term is considered, we simplify and add the nonlinear source term only when $i = 1$ in Eq. (\ref{CKIter-ref-matrix}).

\begin{align}
    \mathcal{D}_{lp}^1 |J| M_{kl}
      = - |J| \Theta^{-1}_{ed} \mathcal{D}_{lq}^{0} B_{pq}^{d,t_n} K^e_{lk}
      + |J| \int_{\mathcal{E}_3} s_p(\underset{-}{q}^{t_n}) \phi_k \text{d} V
    \label{CKIter-ref-source},
\end{align}
where $u_q^{t_n} = \mathcal{D}_{lq}^0 \phi_l$, with the same definition of $\Theta^{-1}_{ed}$ as Eq. (\ref{CKIter-ref}). The nonlinear source function $s_p(\underset{-}{u}^{t_n})$ is evaluated on a nodal basis of $u_q^{t_n}$ projected from the modal basis coefficients $\mathcal{D}_{lq}^0$ as presented by \citet{wollherr2018off}.

\subsection{Correction step: time integration and discontinuity handling}
\label{subsubsec:correction_step}

The weak form of Eq. (\ref{nonlinear hyper pdes}) with integration by part looks like
\begin{align}
    \dfrac{\partial}{\partial t} \int_{\mathcal{T}_m} \phi_k U_{lp}(t) \phi_l \text{d} V
     + \int_{\partial \mathcal{T}_m} \phi_k (F_p^d n_d)^{*}
    \text{d} S 
    - \int_{\mathcal{T}_m}
    \dfrac{\partial \phi_k}{\partial x_d} F_p^d 
    \text{d} V 
    \label{Int-by-part} 
    = \int_{\mathcal{T}_m} s_p(U_{lp} \phi_l) \phi_k \text{d} V,
\end{align}
where $ s_p(U_{lp} \phi_l) = (0,0,0,0,0,0,0,0,0,r_{\alpha} )^T$ as in Eq. (\ref{nonlinear hyper pdes}). $n_d$ is the normal vector of the interface $\partial \mathcal{T}_m$. Integrating both sides of the Eq. (\ref{Int-by-part}) in one time step [$t_n$, $t_{n+1}$] yields

\begin{align}
    \int_{\mathcal{T}_m} \phi_k \phi_l [Q^{n+1}_{lp} - U^{n}_{lp}] \text{d}V
     + \int_{\partial \mathcal{T}_m} \phi_k 
    \int_{t_{n}}^{t_{n+1}} (F_p^d n_d)^{*}
    \text{d} \tau
    \text{d} S 
    - \int_{\mathcal{T}_m} 
    \dfrac{\partial \phi_k}{\partial x_d} 
    \int_{t_{n}}^{t_{n+1}} F_p^d
    \text{d} \tau
    \text{d} V \nonumber \\
    = \int_{\mathcal{T}_m} \phi_k
    \int_{t_{n}}^{t_{n+1}} s_p(U_{lp} \phi_l) 
    \text{d} \tau
    \text{d} V.
    \label{time-integrated}
\end{align}

According to Eqs. (\ref{modal dg space}) and (\ref{eq:taylor_expansion}), we estimate the space-time integration in each term of Eq. (\ref{time-integrated}) with $\mathcal{D}_{lp}^i$ derived from the prediction step. 

We expand on the space-time integration term by term in the following.
We start from the second term on the left-hand-side of Eq. (\ref{time-integrated}) when $\partial \mathcal{T}_m$ is on the element surfaces that are not on the boundaries of the computation domain. The latter case will be addressed in Section \ref{subsec:boundary_conditions}. The interface flux within the computational domain $(F_p^d n_d)^*$ must account for the solution discontinuities on each side of the interface. Strictly speaking, this requires solving the Riemann problem for a nonlinear hyperbolic system \citep{leveque2002finite}. Here we use the local Lax-Friedrich flux $F^{LF}_p$ which has a simple form while preserving numerical stability. Its expression is

\begin{align}
    F^{LF}_p
    &= (F_p^d n_d)^{*}_p \nonumber \\
    &= 
    \dfrac{1}{2} (
    \textcolor{black}{ {F}_p^d (u_p^+)  } 
    + {F}_p^d (u_p^-)) n_d
    + \dfrac{1}{2} C (u_p^- - \textcolor{black} {u_p^+} ),
    \label{rusanov-flux}
\end{align}
where $C$ is the largest eigenvalues of the matrix $B_{pq}^d((\underset{-}{u}^+ + \underset{-}{u}^-)/2)$ in Eq. (\ref{CKIter}). As defined in Eq. (\ref{rusanov-flux}), $F^{LF}_p$ is a nonlinear function of $u_p$ on both sides of $u_p^+$ and $u_p^-$. For the numerical integration, we evaluate $F^{LF}_p$ at the quadrature points in space and time following \citet{uphoff2020flexible} and expand the second term on the left-hand-side of Eq. (\ref{time-integrated}) as

\begin{align}
    \int_{\partial \mathcal{T}_m} \phi_k 
    \int_{t_{n}}^{t_{n+1}} (F_p^d n_d)^{*}
    \text{d} \tau
    \text{d} S \nonumber \\
    = \sum_{i=1}^{N^s}  \beta_i \phi_{k,i} 
    \sum_{z=1}^{N^t} \gamma_z F^{LF}_{lp,z,s}
    |S_f| \Delta t,
    \label{2nd term}
\end{align}
where $\beta_i$ and $\gamma_z$ are weights, respectively, for surface and time integration.

For the third term on the left-hand-side of Eq. (\ref{time-integrated}), we also discretize $F^{d}_p = \mathcal{F}^{d}_{lp}(t) \phi_l(\boldsymbol{x})$ with the same modal basis functions as $u_p$.  We briefly summarize the procedures here and refer to \citet{wollherr2018off} for the detailed formulae. The evaluation of $\mathcal{F}^{d}_{lp}(t)$ follows 3 steps: (1) Project $U_{lp}(t)$ into a nodal basis and obtain the $U^{Node}_{lp}(t)$ coefficients in the nodal basis; (2) Evaluate the coefficients $\mathcal{F}^{d, Node}_{lp}$ in nodal space by substituting $U^{Node}_{lp}(t)$ into the nonlinear function $\mathcal{F}^{d}_{p}(U^{Node}_{lp})$ based on Eq. (\ref{nonlinear damage models}) to Eq. (\ref{cdb-model}); (3) Obtain the coefficients $\mathcal{F}^{d}_{lp}(t)$ in modal space by projecting back from the nodal space coefficients $\mathcal{F}^{d}_{p}(U^{Node}_{lp})$. The third term on the left-hand-side of Eq. (\ref{time-integrated}) then becomes

\begin{align}
    \int_{\mathcal{T}_m} 
    \dfrac{\partial \phi_k}{\partial x_d} 
    \int_{t_{n}}^{t_{n+1}} F_p^d
    \text{d} \tau
    \text{d} V \nonumber \\
    = \int_{t_{n}}^{t_{n+1}} \mathcal{F}^{d}_{lp}(\tau)
    \text{d} \tau
    \int_{\mathcal{T}_m} 
    \dfrac{\partial \phi_k}{\partial x_d} \phi_l
    \text{d} V
    \label{3rd term}.
\end{align}

We employ a similar procedure for the right-hand-side of Eq. (\ref{time-integrated}). We discretize $s_p(t) = S_{lp}(t)\phi_l$ and yield
\begin{align}
    \int_{\mathcal{T}_m} \phi_k
    \int_{t_{n}}^{t_{n+1}} s_p(U_{lp} \phi_l) 
    \text{d} \tau
    \text{d} V \nonumber \\
    = \int_{t_{n}}^{t_{n+1}} S_{lp}(\tau)
    \text{d} \tau
    \int_{\mathcal{T}_m} 
    \phi_k \phi_l
    \text{d} V
    \label{rhs term}.
\end{align}

\subsection{Free surface and absorbing boundary conditions}
\label{subsec:boundary_conditions}

We need to take care of the numerical flux $(F_p^d n_d)^{*}$ in the second term of Eq. (\ref{time-integrated}) when $\partial \mathcal{T}_m$ is defined on two types of boundaries that are important for earthquake simulations: the absorbing boundary and the free-surface boundary. While IVM in Eq. (\ref{ivm-model}) remains isotropic with damage accumulation, CDM in Eq. (\ref{cdb-model}) can introduce stress-induced anisotropic mechanical responses in rocks \citep{hamiel2009brittle}. Such anisotropy inside the bulk materials can be resolved using the local Lax-Friedrich flux in Eq. (\ref{rusanov-flux}) \citep{de2007arbitrary}. In defining the boundary conditions of the simulation domain, we simplify by only considering the nonlinear effects on the isotropic moduli, i.e., the two Lam\'{e} parameters. To achieve this, we retain only the components of $\underset{=}{B}^d = B^d_{pq}$ that correspond to the isotropic effective Lam\'{e} parameters, denoting an approximated matrix as $\underset{=}{B}^{d,eff}$. The expressions for $\underset{=}{B}^{d,eff}$ are \citep{wilcox2010high}:
\begin{align}
    \underset{=}{B}^{1,eff}
    = \begin{bmatrix}
        0 &0 &0 &0 &0 &0 &-1 &0 &0 &0\\
        0 &0 &0 &0 &0 &0 &0 &0 &0 &0\\
        0 &0 &0 &0 &0 &0 &0 &0 &0 &0\\
        0 &0 &0 &0 &0 &0 &0 &-\dfrac{1}{2} &0 &0\\
        0 &0 &0 &0 &0 &0 &0 &0 &0 &0\\
        0 &0 &0 &0 &0 &0 &0 &0 &-\dfrac{1}{2} &0\\
        -\dfrac{\lambda^{eff}+2\mu^{eff}}{\rho} &-\dfrac{\lambda^{eff}}{\rho}   &-\dfrac{\lambda^{eff}}{\rho} &0 &0 &0 &0 &0 &0 &0\\
        0 &0 &0 &-\dfrac{2\mu^{eff}}{\rho} &0 &0 &0 &0 &0 &0\\
        0 &0 &0 &0 &0 &-\dfrac{2\mu^{eff}}{\rho} &0 &0 &0 &0\\
        0   &0   &0 &0 &0 &0 &0 &0 &0 &1
    \end{bmatrix}
    \label{eq:matrix_Beff_1},
\end{align}

\begin{align}
    \underset{=}{B}^{2,eff}
    = \begin{bmatrix}
        0 &0 &0 &0 &0 &0 &0 &0 &0 &0\\
        0 &0 &0 &0 &0 &0 &0 &-1 &0 &0\\
        0 &0 &0 &0 &0 &0 &0 &0 &0 &0\\
        0 &0 &0 &0 &0 &0 &-\dfrac{1}{2} &0 &0 &0\\
        0 &0 &0 &0 &0 &0 &0 &0 &-\dfrac{1}{2} &0\\
        0 &0 &0 &0 &0 &0 &0 &0 &0 &0\\
        0 &0 &0 &-\dfrac{2\mu^{eff}}{\rho} &0 &0 &0 &0 &0 &0\\
        -\dfrac{\lambda^{eff}+2\mu^{eff}}{\rho} &-\dfrac{\lambda^{eff}}{\rho}   &-\dfrac{\lambda^{eff}}{\rho} &0 &0 &0 &0 &0 &0 &0\\
        0 &0 &0 &0 &-\dfrac{2\mu^{eff}}{\rho} &0 &0 &0 &0 &0\\
        0   &0   &0 &0 &0 &0 &0 &0 &0 &1
    \end{bmatrix}
    \label{eq:matrix_Beff_2},
\end{align}

\begin{align}
    \underset{=}{B}^{3,eff}
    = \begin{bmatrix}
        0 &0 &0 &0 &0 &0 &0 &0 &0 &0\\
        0 &0 &0 &0 &0 &0 &0 &0 &0 &0\\
        0 &0 &0 &0 &0 &0 &0 &0 &-1 &\\
        0 &0 &0 &0 &0 &0 &0 &0 &0 &0\\
        0 &0 &0 &0 &0 &0 &0 &-\dfrac{1}{2} &0 &0\\
        0 &0 &0 &0 &0 &0 &-\dfrac{1}{2} &0 &0 &0\\
        0 &0 &0 &0 &0 &-\dfrac{2\mu^{eff}}{\rho} &0 &0 &0 &0\\
        0 &0 &0 &0 &-\dfrac{2\mu^{eff}}{\rho} &0 &0 &0 &0 &0\\
        -\dfrac{\lambda^{eff}}{\rho} &-\dfrac{\lambda^{eff}+2\mu^{eff}}{\rho} &-\dfrac{\lambda^{eff}}{\rho} &0 &0 &0 &0 &0 &0 &0\\
        0   &0   &0 &0 &0 &0 &0 &0 &0 &1
    \end{bmatrix}
    \label{eq:matrix_Beff_3}.
\end{align}

The effective Lam\'{e} parameters for IVM are

\begin{equation}
  \begin{cases}
      \lambda^{eff} = (1-\alpha) \lambda_0  \\
      \mu^{eff} = (1-\alpha) \mu_0
    \end{cases}.
    \label{effective-ivm}
\end{equation}

The effective Lam\'{e} parameters for CDM are

\begin{equation}
  \begin{cases}
      \lambda^{eff} = \lambda_0 - \alpha \gamma_r \epsilon / \sqrt{I_2} \\
      \mu^{eff} = \mu_0 - \alpha \xi_0 \gamma_r -  0.5 \alpha \gamma_r \xi
    \end{cases},
    \label{effective-cdb}
\end{equation}
where $\epsilon = (\varepsilon_{xx}+\varepsilon_{yy}+\varepsilon_{zz})/3$.

We compute the numerical fluxes $(F_p^d n_d)^{*}$ on both the absorbing boundary and the free-surface boundary based on the solutions of the Riemann problem with an upwind method using the approximate effective matrix $\underset{=}{B}^{d,eff}$ defined in Eq. (\ref{eq:matrix_Beff_1}) to (\ref{eq:matrix_Beff_3}). We assume that the outgoing waves at the element interface are only influenced by the state in the element that the interface belongs to; the incoming waves at the element interface are only influenced by the state in the neighboring element.

To compute the upwind flux, we diagonalize matrix $\underset{=}{B}^{1,eff} = \underset{=}{R} \underset{=}{\Lambda} \underset{=}{R}^{-1}$, where $\underset{=}{\Lambda} = \diag{-c_p^{eff},-c_s^{eff},-c_s^{eff},0,0,0,c_s^{eff},c_s^{eff},c_p^{eff},0}$, $c_p^{eff} = \sqrt{(\lambda_{eff} + 2\mu_{eff}) / \rho}$, $c_p^{eff} = \sqrt{\mu_{eff} / \rho}$, and
\begin{align}
    \underset{=}{R}
    = \begin{bmatrix}
        1 &0 &0 &-\dfrac{\lambda^{eff}}{\lambda^{eff} + 2 \mu^{eff}} &0 &-\dfrac{\lambda^{eff}}{\lambda^{eff} + 2 \mu^{eff}} &0 &0 &-1 
        &0\\
        0 &0 &0 &0 &0 &1 &0 &0 &0 &0\\
        0 &0 &0 &1 &0 &0 &0 &0 &0 &0\\
        0 &\dfrac{1}{2} &0 &0 &0 &0 &0 &-\dfrac{1}{2} &0 
        &0\\
        0   &0   &0 &0 &1 &0 &0 &0 &0 &0\\
        0   &0   &\dfrac{1}{2} &0 &0 &0 &-\dfrac{1}{2} &0 &0 
        &0\\
        c_p^{eff}   &0   &0 &0 &0 &0 &0 &0 &c_p^{eff} &0\\
        0 &c_s^{eff} &0 &0 &0 &0 &0 &c_s^{eff} &0 &0\\
        0   &0   &c_s^{eff} &0 &0 &0 &c_s^{eff} &0 &0 &0\\
        0   &0   &0 &0 &0 &0 &0 &0 &0 &1
    \end{bmatrix}
    \label{matrix-R},
\end{align}
where the last column results from the extra zero eigenvalues due to the introduction of the damage variable.

For the absorbing boundaries, we use the same method as \citet{dumbser2006arbitrary}.

\begin{align}
    F^{abs}_p
    &= (F_p^d n_d)^{*}_p \nonumber \\
    &= T_{pq} B^{1,eff,+}_{qr} T_{rs}^{-1} q_s
    \label{absorbing-flux},
\end{align}
where $\underset{=}{B}^{1,eff,+} = \underset{=}{R} \underset{=}{\Lambda^+} \underset{=}{R}^{-1}$. $\underset{=}{\Lambda^+} = \diag{0,0,0,0,0,0,c_s^{eff},c_s^{eff},c_p^{eff},0}$ only keeps the positive terms in $\underset{=}{\Lambda}$. $T_{pq}^{-1}$ is the rotation matrix that operates on the vector of the conservative variables $u_s$, rotating the quantities to the face-aligned coordinate system.

\begin{align}
    \underset{=}{T}
    = \begin{bmatrix}
        n_x^2 &s_x^2 &t_x^2 
        &2 n_x s_x &2 s_x t_x &2 t_x n_x 
        &0 &0 &0 &0\\
        n_y^2 &s_y^2 &t_y^2 
        &2 n_y s_y &2 s_y t_y &2 t_y n_y 
        &0 &0 &0 &0\\
        n_z^2 &s_z^2 &t_z^2 
        &2 n_z s_z &2 s_z t_z &2 t_z n_z 
        &0 &0 &0 &0\\
        n_x n_y &s_x s_y &t_x t_y 
        &n_x s_y + n_y s_x &t_x s_y + t_y s_x &n_x t_y + n_y t_x
        &0 &0 &0 &0\\
        n_z n_y &s_z s_y &t_z t_y 
        &n_z s_y + n_y s_z &t_z s_y + t_y s_z &n_z t_y + n_y t_z
        &0 &0 &0 &0\\
        n_x n_z &s_x s_z &t_x t_z 
        &n_x s_z + n_z s_x &t_x s_z + t_z s_x &n_x t_z + n_z t_x
        &0 &0 &0 &0\\
        0   &0   &0 
        &0  &0   &0 
        &n_x &s_x &t_x &0\\
        0   &0   &0 
        &0  &0   &0 
        &n_y &s_y &t_y &0\\
        0   &0   &0 
        &0  &0   &0 
        &n_z &s_z &t_z &0\\
        0   &0   &0 &0 &0 &0 
        &0 &0 &0 &1
    \end{bmatrix}
    \label{matrix-T}.
\end{align}

For the free-surface boundaries, we first rotate $u_q$ to the face-aligned coordinate system as $u^n_p = T_{rs}^{-1} u_s$. We then derive the constraints to the conservative variables $u^b_q$ on the boundary face from an upwind flux below in a similar way as \citet{uphoff2020flexible}.

\begin{align}
    \underset{-}{u}^b
    &= \underset{-}{u}^- + \omega_1  \underset{-}{r}^1 + \omega_2 \underset{-}{r}^2 +  \omega_3 \underset{-}{r}^3 \nonumber \\
    &= \underset{-}{u}^- + \omega_1 \begin{pmatrix}
        1 \\
        0 \\ 0 \\ 0 \\ 0 \\ 0 \\
        c_p^{eff} \\
        0 \\ 0 \\ 0
    \end{pmatrix}
    + \omega_2 \begin{pmatrix}
        0 \\
        0 \\ 0 \\ 1/2 \\ 0 \\ 0 \\
        0 \\
        c_s^{eff} \\ 0 \\ 0
    \end{pmatrix}
    + \omega_3 \begin{pmatrix}
        0 \\
        0 \\ 0 \\ 0 \\ 0 \\ 1/2 \\
        0 \\
        0 \\ c_s^{eff} \\ 0
    \end{pmatrix}
    \label{q_b},
\end{align}
where $\underset{-}{u}^-$ is the projection of solutions in the local element on the free surface; $\underset{-}{r}^1$ is the column in $\underset{=}{R}$ that corresponds to -$c_p^{eff}$ in $\underset{=}{\Lambda}$; $\underset{-}{r}^2$ and $\underset{-}{r}^3$ are the two columns in $\underset{=}{R}$ that correspond to -$c_s^{eff}$ in $\underset{=}{\Lambda}$. $\omega_1$, $\omega_2$ and $\omega_3$ are unknowns to be constrained from the free-surface boundary conditions, which we will further define below.

We derive from $u^b_p$ the face-aligned boundary stress $u_p^{\sigma,b} = (\sigma_{xx},\sigma_{yy},\sigma_{zz},\sigma_{xy},\sigma_{yz},\sigma_{zx}, v_x, v_y, v_z, \alpha)^T$, where
\begin{align}
    u_p^{\sigma,b} 
    &= C_{pq} u^b_q \nonumber \\
    &= \begin{bmatrix}
        \lambda^{eff} + 2 \mu^{eff} &\lambda^{eff} &\lambda^{eff} & &0 &0 &\\
        \lambda^{eff} &\lambda^{eff} + 2 \mu^{eff} &\lambda^{eff} &0 &0 &1 &\\
        \lambda^{eff} &\lambda^{eff} &\lambda^{eff} + 2 \mu^{eff} &0 &0 &0 &\underset{=}{0}\\
        0 &0 &0 &2 \mu^{eff} &0 &0 &\\
        0 &0 &0 &0 &2 \mu^{eff} &0 &\\
        0 &0 &0 &0 &0 &2 \mu^{eff} &\\
          & &\underset{-}{0} & & & &\underset{=}{I}
    \end{bmatrix}  \underset{-}{u}^{b} 
    \label{matrix-C},
\end{align}
where $\underset{=}{I}$ is a 4 by 4 identity matrix, while $\underset{=}{0}$ and $\underset{-}{0}$ are, respectively, the zero matrix and zero vector that complete the matrix $C_{pq}$.

On the free surface, $\sigma_{xx}$, $\sigma_{xy}$ and $\sigma_{zx}$ in $u_p^{\sigma,b}$ should be zero. With these three more constraints, we solve the unknowns $\omega_1$, $\omega_2$ and $\omega_3$ in Eq. (\ref{q_b}). We can substitute these unknowns back in Eq. (\ref{q_b}) to obtain $u_q^b$ in the face-aligned coordinate system. We finally compute the boundary flux with $u_q^b$ as below.

\begin{align}
    F^{free}_p
    &= (F_p^d n_d)^{*}_p \nonumber \\
    &= T_{pq} B^{1,eff}_{qr} u_r^b.
    \label{freeSurface-flux}
\end{align}

\section{Frequency components of the ground motion recorded at different stations inside the Kathmandu Valley}
\label{sec:app_more_stations}

\begin{figure}[hptb]
    \centering
    \includegraphics[width=0.9\columnwidth]{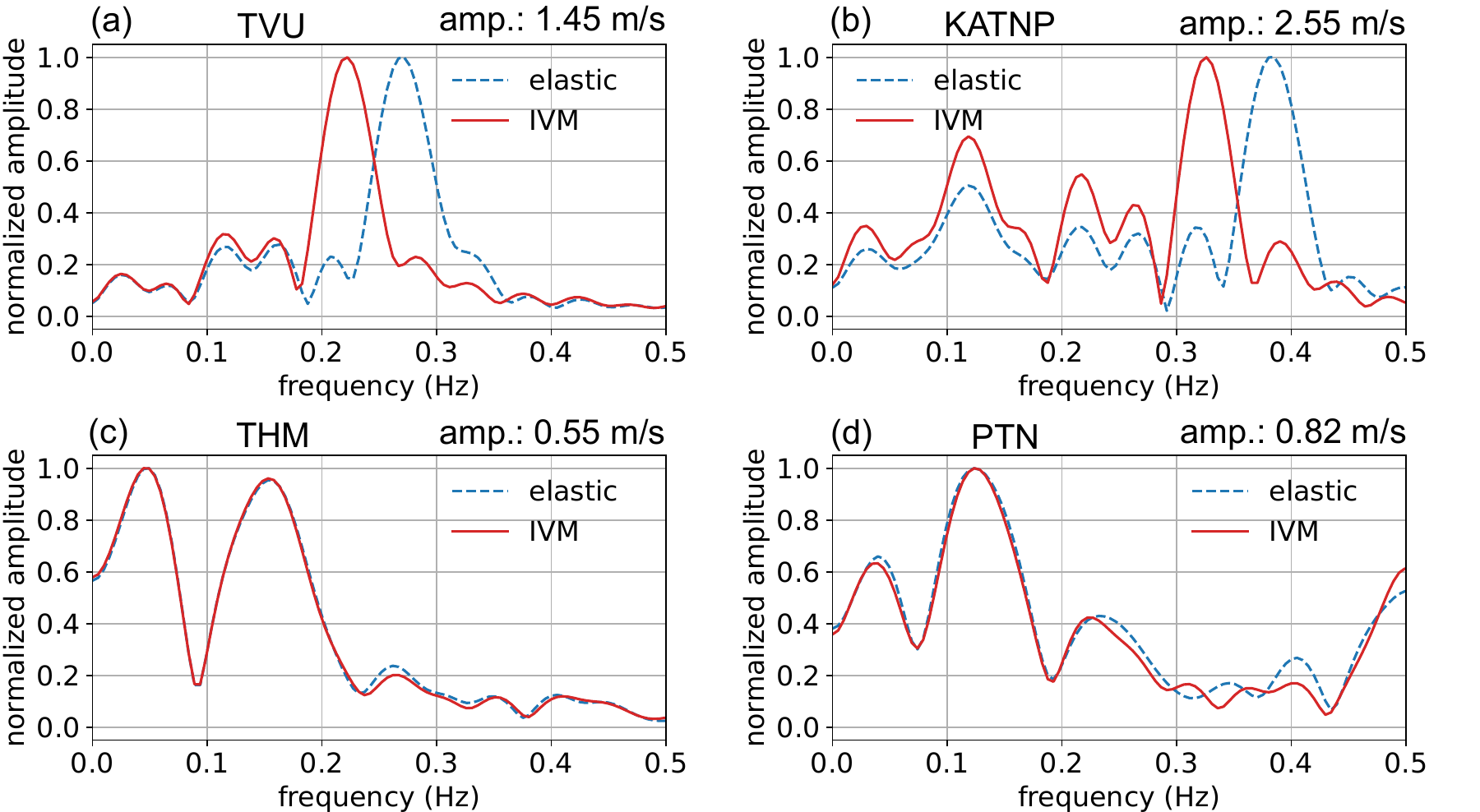}
    \caption{\small Normalized frequency spectra of the upward-downward velocity time series recorded between 20 s and 50 s for simulations employing elastic model (the dashed blue curve) and IVM (the solid red curve) at 4 stations: (a) TVU, (b) KATNP, (c) THM, and (d) PTN. We provide peak magnitudes of the velocity vector at the four stations on the top right of each subfigure.}
    \label{fig:kathmandu_freq_more}
\end{figure}

This section provides supporting information for reproducing the low-frequency enhancement observed in ground motions from our simulations using the nonlinear damage model, IVM. We present comparisons between the frequency components of the velocity time series predicted by the elastic model and the IVM at four additional strong motion stations within the Kathmandu Valley, as shown in Fig. \ref{fig:kathmandu_freq_more}. Among the four listed stations, we find a prominent low-frequency enhancement in the simulations with IVM between 0.2 Hz and 0.35 Hz at station TVU and between 0.3 Hz and 0.45 Hz at station KATNP. In contrast, the frequency spectra at stations THM and PTN show negligible differences between simulations using the linear elastic model and those with IVM. peak ground velocity (PGV) is strongly correlated with the prominence of the low-frequency enhancement. Specifically, the PGV at station TVU and KATNP is approximately twice and four times higher, respectively, than at station THM. The PGV at the station PTN is approximately 60\% larger than that at station THM. With such an intermediate PGV value, only a minor low-frequency enhancement between 0.25 Hz and 0.4 Hz is observed at station PTN. The more prominent low-frequency enhancement associated with larger PGV is attributed to the stronger reduction in co-seismic moduli in regions with high PGV values.

\bibliography{references}

%
%
%
%
%

\end{document}